\documentclass[longauth]{aa}   

\usepackage{graphicx}
\usepackage{txfonts}
\usepackage[dvipsnames]{xcolor}
\usepackage{lipsum}
\usepackage{sidecap}

\usepackage{mathtools}
\usepackage{dirtytalk}
\usepackage{enumitem}
\usepackage{amsmath}
\usepackage{multirow,multicol}
\usepackage{microtype}
\microtypesetup{protrusion=true,expansion=true}
\usepackage{hyperref}
\hypersetup{
    colorlinks=false,
    linkcolor=blue,
    filecolor=magenta,      
    urlcolor=cyan
    }
\usepackage{subcaption}         
\usepackage{lscape}             
\usepackage{placeins}           

\newcommand{\appropto}{\mathrel{\vcenter{
  \offinterlineskip\halign{\hfil$##$\cr
    \propto\cr\noalign{\kern2pt}\sim\cr\noalign{\kern-2pt}}}}}

\begin{document}

   \title{Prevalence of radio Active Galactic Nuclei in galaxy groups \\since $z=3.5$ in the COSMOS-Web field}


%
%
%

   \author{F. Ubertosi\inst{1,2}\fnmsep\thanks{Contact: \texttt{francesco.ubertosi2@unibo.it}}
          \and
        I. Delvecchio\inst{3} 
        \and
        I. Prandoni\inst{2,13}
        \and
        G. Toni\inst{1,3,4}
        \and
        G. Zamorani\inst{3}
        \and
        E. Daddi\inst{5}
        \and
        M. Gitti\inst{1,2}
        \and
        M. Sargent\inst{6}
        \and
        A. Traina\inst{3}
        \and 
        C. Spingola\inst{2}
        \and
        F. Vito\inst{3}
        \and
        L. Moscardini\inst{1}
        \and
        G. Gozaliasl\inst{8,9}
        \and
        D. Kakkad\inst{10,11}
        \and
        W. Wang\inst{12}
        \and
        S. Sanjaripour\inst{12}
        \and
        Z. Ghaffari\inst{14,15}
        \and
        M. N. Isla Llave\inst{1,3}
        \and
        R. M. Samir\inst{28}
        \and
        G. Peluso\inst{3}
        \and
        M. Maturi\inst{4,29}
        \and 
        V. Smolcic\inst{7}
        \and
        C. Casey\inst{18,19,20}
        \and
        J. Kartaltepe\inst{16}
        \and
        H. J. McCracken\inst{17}
        \and
        J. Rhodes\inst{23}
        \and
        B. Robertson\inst{24}
        \and
        M. Franco\inst{5,19}
        \and
        D. Liu\inst{25}
        \and
        M. Shuntov\inst{20,21,22}
        \and
        H. Akins\inst{19}
        \and
        O. Ilbert\inst{26}
        \and
        A. A. Khostovan\inst{16,27}
        } 
        
\institute{Dipartimento di Fisica e Astronomia, Università di Bologna, via Gobetti 93/2, I-40129 Bologna, Italy,
         \and
             Istituto Nazionale di Astrofisica - Istituto di Radioastronomia (IRA), via Gobetti 101, I-40129 Bologna, Italy
             \and
        INAF – Osservatorio di Astrofisica e Scienza dello Spazio di Bologna, Via Gobetti 93/3, I-40129 Bologna, Italy
             \and
             Zentrum für Astronomie, Universität Heidelberg, Philosophenweg 12, 69120 Heidelberg, Germany
             \and
             Université Paris-Saclay, Université Paris Cité, CEA, CNRS, AIM, 91191, Gif-sur-Yvette, France
             \and
             Institute of Physics, Laboratory of Astrophysics, \'Ecole Polytechnique F\'ed\'erale de Lausanne (EPFL), Observatoire de Sauverny, Versoix CH-1290, Switzerland
             \and
             Department of Physics, University of Zagreb, Bijeniˇcka cesta 32, Zagreb, Croatia
             \and
             Department of Computer Science, Aalto University, PO Box 15400, Espoo, FI-00076, Finland
             \and
             Department of Physics, University of Helsinki, P. O. Box 64, FI-00014 Helsinki, Finland
             \and
             Centre for Astrophysics Research, Department of Physics, Astronomy and Mathematics, University of Hertfordshire, Hatfield, AL10 9AB, UK
             \and
             Space Telescope Science Institute, 3700 San Martin Drive, Baltimore, 21218 MD, USA
             \and
             Caltech/IPAC, 1200 E. California Blvd. Pasadena, CA 91125, USA
             \and
             Department of Astronomy, University of Cape Town, 7701 Rondebosch, Cape Town, South Africa
             \and
             INAF Osservatorio Astronomico di Trieste, Via G.B. Tiepolo 11 e via Bazzoni 2. 34100, Trieste, Italy
             \and
             IFPU: Institute for Fundamental Physics of the Universe, Via Beirut, 2, 34151, Italy
             \and
             Laboratory for Multiwavelength Astrophysics, School of Physics and Astronomy, Rochester Institute of Technology, 84 Lomb Memorial Drive, Rochester, NY 14623, USA
             \and
             Institut d'Astrophysique de Paris, UMR 7095, CNRS, and Sorbonne Université, 98 bis boulevard Arago, 75014 Paris, France
             \and
             Department of Physics, University of California, Santa Barbara, Santa Barbara, CA 93106, USA
             \and
             The University of Texas at Austin, 2515 Speedway Blvd Stop C1400, Austin, TX 78712, USA
             \and
             Cosmic Dawn Center (DAWN), Denmark
             \and
             Niels Bohr Institute, University of Copenhagen, Jagtvej 128, 2200 Copenhagen, Denmark
             \and
             Département d'Astronomie, Université de Genève, 24 rue du Général-Dufour, 1211 Genève 4, Switzerland
             \and
             Jet Propulsion Laboratory, California Institute of Technology, 4800 Oak Grove Drive, Pasadena, CA 91109, USA
             \and
             Department of Astronomy and Astrophysics, University of California, Santa Cruz, 1156 High Street, Santa Cruz, CA 95064, USA
             \and
             Purple Mountain Observatory, Chinese Academy of Sciences, 10 Yuanhua Road, Nanjing 210023, PR China
             \and
             Aix Marseille Univ, CNRS, LAM, Laboratoire d'Astrophysique de Marseille, Marseille, France
             \and
             Department of Physics and Astronomy, University of Kentucky, 505 Rose Street, Lexington, KY 40506, USA
             \and
             Department of Astronomy, National Research Institute of Astronomy and Geophysics (NRIAG), Cairo, 11421, Egypt
             \and
             Institute for Theoretical Physics, Philosophenweg 16, D-69120 Heidelberg, Germany
            }
   \date{}

 
  \abstract
{}
{Radio active galactic nuclei (AGN) are key tracers of kinetic feedback from supermassive black holes. Although the dependence of their triggering on host stellar mass is well known, the {impact} of the large-scale environment across cosmic time is poorly understood. We investigate the role of the environment 
in triggering radio AGN activity, while also studying the dependence on stellar mass ($10\leq\log M_{\ast}\,{\rm [M_{\odot}]}\leq11.4$), radio luminosity ($22\leq\log L_{1.4}\,{\rm [W\,Hz^{-1}]}\leq26$), and redshift ($0.1\leq z \leq 3.5$).}
{We combine the VLA-COSMOS 3\,GHz Large Project with the COSMOS-Web galaxy and group catalogs, selecting radio-excess AGN and building radio luminosity functions (RLF) and radio AGN fractions separately for sources in galaxy groups (124 radio AGN) and in the field (405 radio AGN), out of the total of 20203 galaxies and 1678 galaxy groups. We fit the RLF to derive the cosmic evolution of radio AGN in different environments, and we provide a new parametric description of the radio AGN fractions as a function of environment, $M_{\ast}$, $L_{1.4}$, and $z$.}
{Besides confirming the previously known positive correlation of radio AGN fraction with host $M_{\ast}$ in both galaxy groups and in the field, more importantly we find that the activation of a radio AGN in galaxy groups is enhanced relative to the field by a factor of $\simeq$3--30 in space density and of $\simeq$2 in radio AGN fraction at fixed stellar mass; the enhancement is also a function of $M_{\ast}$, $L_{1.4}$, and $z$. Furthermore, in all redshift bins, the radio AGN fractions peak within $r \leq 0.2R_{200}$. Overall, dense environments enhance radio AGN triggering out to at least $z \sim 3.5$. We suggest the existence of a possible transition in the dominant triggering mechanism of radio AGN activity around $1\leq z \leq 1.5$, potentially related to the intragroup medium in the core of groups at lower redshift and to merger and structure growth at earlier cosmic times.}
{}


   \maketitle

\nolinenumbers

\section{Introduction}

Active galactic nuclei (AGN) represent a key phase of the evolution of supermassive black holes (SMBHs), during which accretion onto the SMBH is followed by ejection of energy into the hosting galaxy and its environment. This energy deposition -- commonly referred to as AGN feedback -- can regulate or quench star formation (SF) and heat the intracluster and intragroup medium (e.g., \citealt{Croton2006, Bower2006, Fabian2012, Kormendy2013, Hardcastle2020}). AGN feedback manifests in two modes: a radiative (or quasar) mode, in which radiation and winds influence the surrounding medium (on scales of up to $10$s of kpc), and a kinetic (or jet) mode, in which jets of relativistic plasma transport mechanical energy to large distances (even $>100\,{\rm kpc}$ from the host galaxy; e.g., \citealt{Merloni2007,McNamara2007,Heckman2014,Hardcastle2020}). The latter manifests through the jet synchrotron emission in the radio band -- for this reason, \say{radio}\footnote{In this Section, we generally refer to radio AGN as the active SMBHs whose jets produce a detectable radio emission.} AGN can be used as tracers of kinetic feedback (e.g., \citealt{Magliocchetti2018b,Igo2025}).

Constraining which factors trigger radio AGN activity is therefore vital for understanding galaxy evolution. Previous studies have established that the occurrence of radio AGN strongly increases with host stellar mass $M_{\ast}$ (e.g., \citealt{Best2005,MauchSadler2007,Sabater2013,Sabater2019,Delvecchio2022,Wang2024}), while at fixed stellar mass it decreases with increasing radio luminosity (e.g., \citealt{Best2005,Wang2024}). The redshift evolution of the radio AGN fraction remains less clear. For instance, \citet{Kondapally2025} and \citet{Ceraj2018} reported little or no evolution with redshift; however, the former study was limited to $z \lesssim 1.5$, while the latter extended to $z\sim6$ but focused on high-luminosity, X-ray detected radio-excess AGN. In contrast, \citet{Wang2024} found that the radio AGN fractions increase up to $z\sim4$, both in quiescent (QG) and star-forming (SFG) galaxies. 
\\\indent Beyond host-galaxy properties, the effects of the large-scale environment in which a galaxy resides has been also investigated as a potential driver of radio AGN activity (see e.g., \citealt{Maglioccheetti2022} for a review). In the local Universe ($0.03 \leq z \leq 0.1$), \citet{Sabater2013} found that, at fixed stellar mass and for 1.4~GHz radio luminosity $L_{1.4} \geq 10^{23}\,{\rm W~Hz^{-1}}$, the radio AGN fraction increases with local galaxy density. At higher redshift, \citet{Hashiguchi2023} showed that the fraction of radio AGN with $L_{1.4} \geq 10^{24}\,{\rm W~Hz^{-1}}$ is higher by a factor of $\sim2$ in galaxy clusters compared to the field up to $z\sim1.4$. Similarly, \citet{Mo2020} studied massive clusters at $0.7 \leq z \leq 1.5$ and found that the radio AGN fraction (for $L_{1.4} \geq 10^{25}\,{\rm W~Hz^{-1}}$) increases with stellar mass in both environments, with cluster values being approximately twice those in the field at fixed stellar mass. Analogous results were found by \citet{Smolcic2011} for radio AGN in X-ray selected galaxy groups at $0.1\leq z\leq1.3$. Recently, \citet{Shah2025} showed that the fraction of AGN between $2\leq z \leq 4$ (identified with a combination of multiple selection criteria, from radio, to mid-infrared, to X-ray wavelengths) increases with the local overdensity of galaxy, and displays a clear excess relative to the field for AGN hosted especially in massive proto-clusters with $M\geq 10^{14}$~M$_{\odot}$ (see also \citealt{Wylezalek2013}. Furthermore, \citet{Wang2025} studied four radio AGN at $z\sim3.5$ and found the ubiquitous presence of gas-rich companion galaxies around them. More recently, \citet{Chapman2026} reported evidence for an over-abundance of radio AGN in a proto-cluster at $z\sim4.3$.

The environmental enhancement of radio AGN activity is, however, not uniform across a cluster or group: radio AGN fractions show enhancements of a factor of 2 - 4 within $r\leq 0.2\times R_{200}$\footnote{The radius within which the average density is 200$\times$ larger than the critical density of the Universe at the redshift of the structure.} (depending on the host stellar mass, \citealt{Best2007,Smolcic2011,Magliocchetti2018b,Hashiguchi2023,deVos2024}). This central enhancement is qualitatively consistent with a scenario in which the hot, dense intracluster medium in cluster cores provides an abundant reservoir of gas that can cool and fuel SMBH accretion (e.g., \citealt{Best2007}). 

In this work, we present a multi-parametric study of radio AGN triggering in the COSMOS field \citep{Scoville2007}. We combine the VLA-COSMOS 3 GHz Large Project \citep{Smolcic2017_continuumdata}, which provides deep radio observations, with photometric and spectroscopic redshifts, stellar masses, and a galaxy groups catalog \citep{Toni2025} built exploiting the recent JWST/NIRCam coverage of COSMOS-Web \citep{Casey2023}. We quantify the role of the large-scale environment -- in terms of membership to a group or to the field, and in terms of projected distance from the center of the group -- in promoting the activation of a radio AGN, while simultaneously also investigating the stellar mass, radio luminosity, and redshift dependence.

The outline of this paper is as follows: we present the catalogs in Sect.~\ref{sec:catalogs}, with details on the procedure used to cross-match them in Sect.~\ref{subsec:createcat}. In Sect.~\ref{sec:methods} we outline the procedure used to build the radio luminosity function of our sources. In Sect.~\ref{sec:analysis} we analyze our results by fitting the luminosity functions (Sect.~\ref{subsub:fitRLF}), by investigating the radio AGN fractions as a function of stellar mass, redshift, radio luminosity, and environment (membership to an overdensity, Sect.~\ref{subsec:fractions}, and its radial dependence, Sect.~\ref{subsec:AGNfraction-vsradius}), and then by fitting the radio AGN fractions (Sect.~\ref{subsec:fitfractions}). We discuss our results in Sect.~\ref{sec:disc}, focusing on the role of environment in triggering a radio AGN (Sect.~\ref{subsec:disc-overdensities}) and on the feedback output of radio AGN in different environments (Sect.~\ref{subsec:intRLF}). We summarize our conclusions in Sect.~\ref{sec:summary}.
We assume a $\Lambda$CDM Cosmology with $\Omega_{\rm M} = 0.3$, $\Omega_{\rm \Lambda} = 0.7$ and $H_{0} = 70\,{\rm km\,s^{-1}\,Mpc^{-1}}$. Magnitudes are given in AB system \citep{Oke1974}. We define the radio spectral index $\alpha$ as $S_{\nu} \propto \nu^{-\alpha}$, where $S_{\nu}$ is the flux density at frequency $\nu$.

\begin{figure}[htp!]
    \centering
        \includegraphics[width=\linewidth]{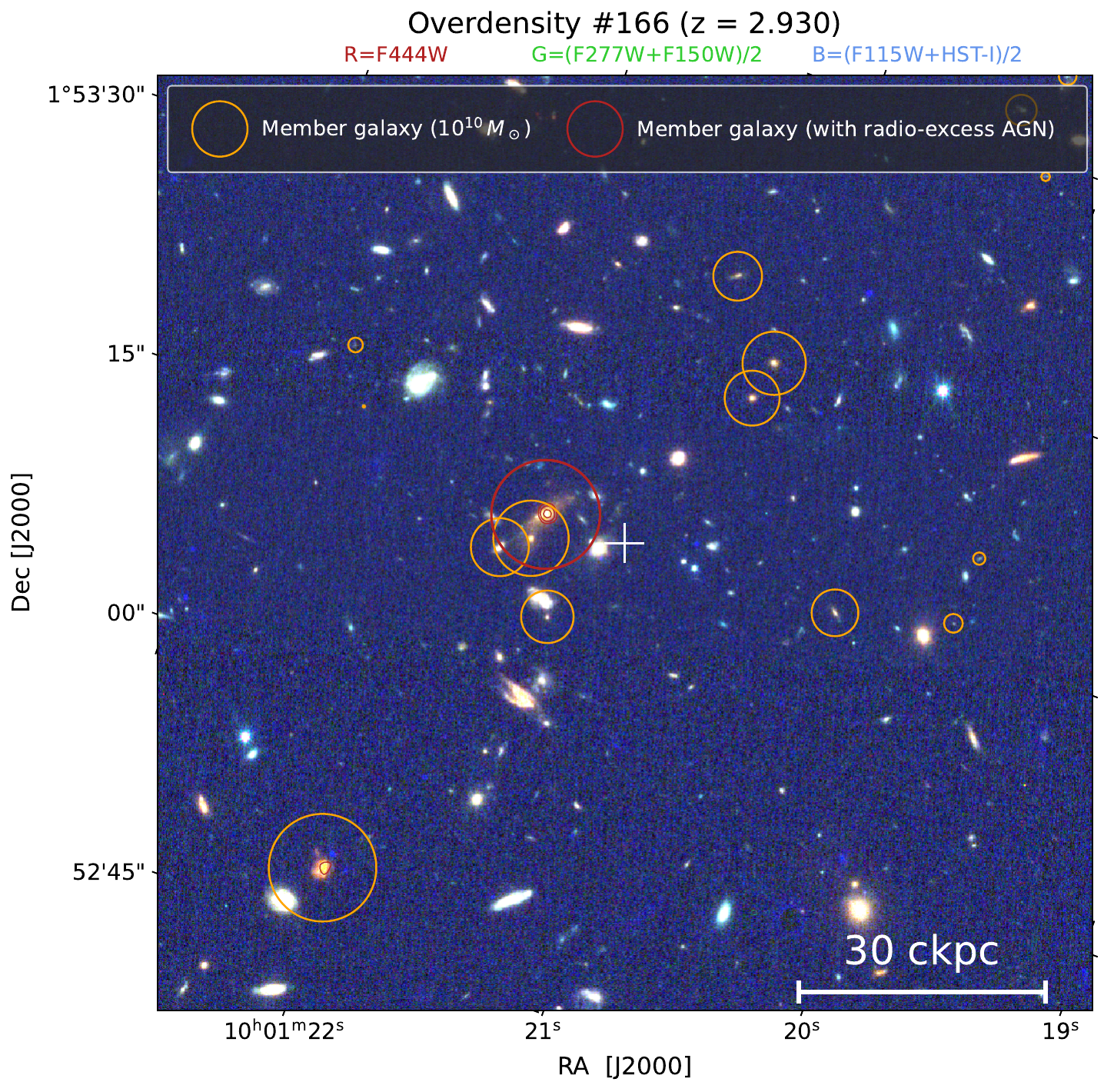}
        \caption{Example of a high-$z$ galaxy group hosting a radio AGN. The figure shows the COSMOS-Web galaxy group \#166 at $z = 2.93$ (M$_{200} = 4.4\times10^{13}$~M$_{\odot}$). Red corresponds to JWST/NIRCam F444W; green corresponds to the average of JWST/NIRCam F277W and F150W; and blue corresponds to the average of JWST/NIRCam F115W \citep{Franco2025} and HST/ACS $I$-band data \citep{Koekemoer2007}. The white plus sign ($+$) denotes the center of the galaxy group from the AMICO detection algorithm \citep{Toni2025}. Red contours (drawn at $5\times, 10\times,...$ the rms noise of $2.3\,\mu$Jy~beam$^{-1}$) are from the VLA-COSMOS 3 GHz survey \citep{Smolcic2017_continuumdata}. Member galaxies (with \texttt{ASSOC\_PROB}$>50\%$) from the catalog of \citet{Toni2025} are identified by orange circles, while members identified as radio-excess AGN from the VLA-COSMOS 3~GHz data (Sect.~\ref{subsec:cosmosvla} \& \ref{subsec:createcat}) are highlighted with red circles. The circle radii are scaled relative to the host $\log M_\ast$, and the labeled circles at the top of the figure are representative of $\log M_\ast {[M_{\odot}]}= 10$; for clarity, members with $M_{\ast}\leq10^{10}$~M$_{\odot}$ -- that is, below the cut imposed for our samples, see Sect.~\ref{subsec:createcat} -- are also marked with circles. 
        }
        \label{fig:multicolor}
\end{figure}

\section{The catalogs: selection and cross-matching}\label{sec:catalogs}

\subsection{COSMOS-Web galaxy catalog}\label{subsec:cosmoswebb}
The COSMOS-Web survey is a JWST program \citep{Casey2023} targeting 0.54 deg$^{2}$ of the COSMOS field, using four NIRCam filters (F115W, F150W, F277W, and F444W) \citep{Franco2025}, and additionally with parallel JWST/MIRI imaging at 7.7$\mu$m over an inner region of 0.2 deg$^2$ \citep{Harish2025}. The survey detected more than 784000 sources, forming the COSMOS2025 catalog; for these sources, data from the COSMOS-Web survey were combined in \citet{Shuntov2025} with additional ground- and space-based data to provide photometric redshifts, morphologies, and physical parameters\footnote{Derived from the template-fitting code \texttt{LePHARE} \citep{Arnouts2002}, with stellar population models selected from \citet{Bruzual2003} as in \citet{Ilbert2015} -- see section 6.1 in \citet{Shuntov2025}.}. Measurements of $M_{\ast}$ were obtained via spectral energy distribution (SED) fitting of the COSMOS-Web ground and space-based photometry, taking the median value of the likelihood distribution for each source.
We used the COSMOS2025 catalog to obtain information on $z$, $M_{\ast}$, and star-formation properties of the objects analyzed in this work. From the catalog, we consider only sources with \texttt{warn\_flag = 0} and with \texttt{flag\_star\_hsc = 0} (see table 2 in \citealt{Shuntov2025} for details), obtaining a catalog of 609010 objects (78\% of the original catalog). We classify galaxies as QG or SFG based on the color-color diagram NUV-r-J (e.g., \citealt{Ilbert2013}), as used in the mass-function of COSMOS-Web galaxies \citep{Shuntov2025}. If available and reliable ($>$95\% confidence level), spectroscopic redshifts from \citet{Khostovan2026} were prioritized over photometric ones (3.3\% of the sources).


\subsection{VLA 3 GHz radio AGN catalog}\label{subsec:cosmosvla}
The 3 GHz observations of the COSMOS field made with the Karl G. Jansky Very Large Array (VLA) provided a mosaic of 2.6~deg$^{2}$ with a median rms noise of 2.3 $\mu$Jy~beam$^{-1}$ at an angular resolution of 0.75$\arcsec$ \citep{Smolcic2017_continuumdata}. From this mosaic, a catalog of 10830 radio sources at a S/N$>$5 was extracted \citep{Smolcic2017_continuumdata,Smolcic2017_multilcounterpart,Smolcic2017_cosmicevo}. We consider this dataset -- one of the deepest and created with one of the highest angular resolutions to date -- to build our sample of radio AGN in the COSMOS-Web field. Specifically, the VLA-COSMOS 3 GHz catalog was first cross-matched with the COSMOS-Web catalog (Sect.~\ref{subsec:cosmoswebb}), providing $z$, $M_{\ast}$, and star-formation rates (SFR), among other properties, for each object. In order to improve the SFR measurements, the VLA 3 GHz catalog was cross-matched (using a 0.5$"$ searching radius) with the de-blended far-IR/sub-millimeter photometric catalog \citep{Jin2018}, which was re-extracted on Ks, Spitzer-MIPS 24 $\mu$m and VLA 3~GHz positional priors (see also \citealt{Liu2018}). If the combined far-IR/sub-mm signal-to-noise from \citet{Jin2018} was larger than five, we adopted their FIR-based SFRs, properly accounting for the updated redshift, if necessary. Then, among the VLA-COSMOS 3~GHz catalog, we select \say{radio-excess} AGN, i.e., with radio luminosity at least 0.44~dex (i.e., $2\sigma$) higher than that expected from star formation alone, following the infrared-radio correlation (IRRC) from \citet{Delvecchio2021} (that has a scatter of 0.22~dex). This approach conservatively selects a sample of 745 radio-excess AGN within COSMOS-Web, with 3~GHz flux densities ranging between $\simeq10\,\mu$Jy and $10^{4}\,\mu$Jy. In the following, we consider the flux-dependent completeness corrections in \citet{Smolcic2017_continuumdata}.


\subsection{COSMOS-Web Group Catalog}\label{subsec:cosmosgroup}
Recently, a catalog of galaxy groups and protocluster cores identified in the COSMOS-Web survey was made available \citep{Toni2024,Toni2025}. This catalog was built by detecting overdensities of galaxies using the Adaptive Matched Identifier of Clustered Objects (AMICO; \citealt{Bellagamba2018}) in the COSMOS2025 catalog \citep{Shuntov2025}, using the galaxy position, photometric redshift, and magnitude in the NIRCam F150W band (down to F150W = 27.3 mag). The resulting catalog includes 1678 galaxy groups between $0.08 \leq z \leq 3.7$
, over an effective area of 0.45~deg$^{2}$ (see \citealt{Toni2025}). In the catalog, each galaxy is associated with a probability (column \texttt{ASSOC\_PROB}) to belong to a specific galaxy group or to the field. Therefore, for a galaxy with e.g., 30\% probability to belong to one group, 10\% probability to belong to a second group, and 60\% probability to belong to the field, there will be three entries in the catalog; with these probabilities, it will more likely belong to the field than to any galaxy group (more details are given in Sect.~\ref{subsec:createcat}). The primary information available for groups is their ID, RA, DEC, redshift, the intrinsic richness $\lambda_{\ast}$ (named \texttt{LAMBDA\_STAR}). The intrinsic richness $\lambda_{\ast}$ is the sum of \texttt{ASSOC\_PROB} for galaxies within R$_{200}^{\rm AMICO}$ of the group model and with ${\rm mag}_{\rm F150W} \leq {\rm mag}_{\rm \ast,F150W} + 1.5$ (where ${\rm mag}_{\rm F150W}$ is the magnitude in the F150W filter and ${\rm mag}_{\rm \ast, F150W}$ is the luminosity function knee magnitude, see \citealt{Toni2025} for details). Additionally, we derived group masses by adopting the methodology presented by \citet{Toni2024} (the application to the full group catalog will be given in Toni et al. in preparation). In short, after applying the same data-cleaning procedure, we fitted a $z$-dependent scaling relation between X-ray-derived masses and $\lambda_{\ast}$ using the COSMOS-Web group sample analyzed in this study \citep{Toni2025}. This provides an updated scaling relation calibrated to the depth and characteristics of the COSMOS-Web dataset. The resulting masses within R$_{200}$ lie in the range $6.5\times10^{12}\,{\rm M}_{\odot}\leq {\rm M}_{200} \leq 10^{14}\,{\rm M}_{\odot}$. We note that the catalog of galaxies in the galaxy group catalog is already filtered for the COSMOS-Web catalog flags listed in Sect.~\ref{subsec:cosmoswebb}. From \citet{Toni2025} we also adopt the corrections for purity and completeness of the galaxy group catalog. These were derived from data-driven mock group catalogs generated with SinFoniA \citep{Maturi2019} and expressed as a function of intrinsic richness $\lambda_{\ast}$ and redshift.



\subsection{Building the catalogs}\label{subsec:createcat}
For the purposes of this work, we combine the catalogs described in Sects.~\ref{subsec:cosmoswebb}, \ref{subsec:cosmosvla}, and \ref{subsec:cosmosgroup} to build the following samples:
\begin{enumerate}[nolistsep]
    \item Galaxies in galaxy groups (\say{GRP}), subdivided into:
    \begin{enumerate}
        \item radio AGN in galaxy groups (\say{${\rm GRP_{rAGN}}$});
        \item non radio AGN galaxies in galaxy groups (\say{${\rm GRP_{GAL}}$}).
    \end{enumerate}
    \item Galaxies in the field (\say{FLD}), subdivided into:
    \begin{enumerate}
        \item radio AGN in the field (\say{${\rm FLD_{rAGN}}$});
        \item non radio AGN galaxies in the field (\say{${\rm FLD_{GAL}}$}).
    \end{enumerate}
\end{enumerate}
We begin by selecting galaxies from the COSMOS-Web group catalog such that each galaxy is associated with a single environment. This is achieved by assigning galaxies to groups when \texttt{ASSOC\_PROB}~$\geq$~0.5, ensuring that they are more likely to belong to an overdensity than to the field (see \citealt{Toni2025}). The remaining galaxies from the COSMOS-Web catalog \citep{Shuntov2025} are instead classified as field galaxies. This procedure yields the two mutually-exclusive samples corresponding to \say{GRP} and \say{FLD} above.
Then, we cross-matched the radio AGN catalog (see Sect.~\ref{subsec:cosmosvla}) with both the group and field galaxy samples using the galaxy identifier. This allows us to identify radio AGN within each environment, yielding the subsamples \say{${\rm GRP_{rAGN}}$} and \say{${\rm FLD_{rAGN}}$}, while the remaining galaxies in each environment constitute the non radio AGN samples, \say{${\rm GRP_{GAL}}$} and \say{${\rm FLD_{GAL}}$}.
Finally, we imposed consistency in stellar mass and redshift between the different catalogs. Specifically, we restricted the catalogs to $10.0\leq\log({\rm M}_{\ast} {\rm[M_{\odot}]})\leq11.4$ and $0.1\leq z\leq 3.5$. The lower cut in stellar masses makes us complete in magnitudes down to the limit of 27.3~mag from the catalog of \citet{Toni2025}. The redshift range was chosen because bright galaxies are underrepresented in COSMOS at $z<0.1$ and to facilitate comparisons with previous studies. An example of a (high-$z$) galaxy group hosting a radio AGN is shown in Fig.~\ref{fig:multicolor}, while a sketch of the subsamples is given in Fig.~\ref{fig:venn}.
\\\indent We note here that the vast redshift range of the galaxy group catalog implies that some of the ${\rm M}_{200} \geq 10^{13.5}\,{\rm M}_{\odot}$ at $z\geq2$ (as the one shown in Fig.~\ref{fig:multicolor}) might also be referred to as protocluster cores (see \citealt{Toni2025}; see also \citealt{Baxter2025} and, for a review, \citealt{Overzier2016}). However, for simplicity, in the following we refer to all the overdensities as \say{galaxy groups}. 

\begin{figure*}[htp!]
    \centering
    \begin{minipage}[c]{0.67\linewidth}
        \includegraphics[width=0.5\linewidth]{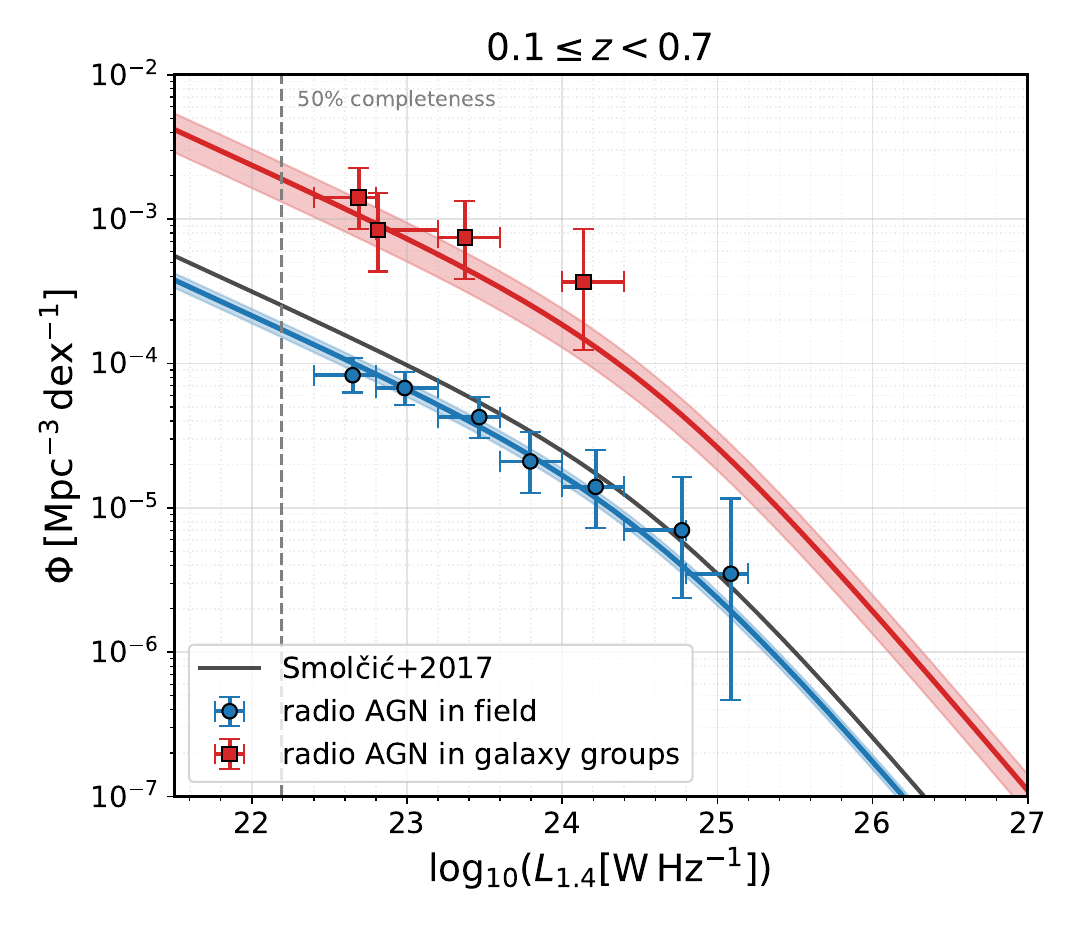}%
        \includegraphics[width=0.5\linewidth]{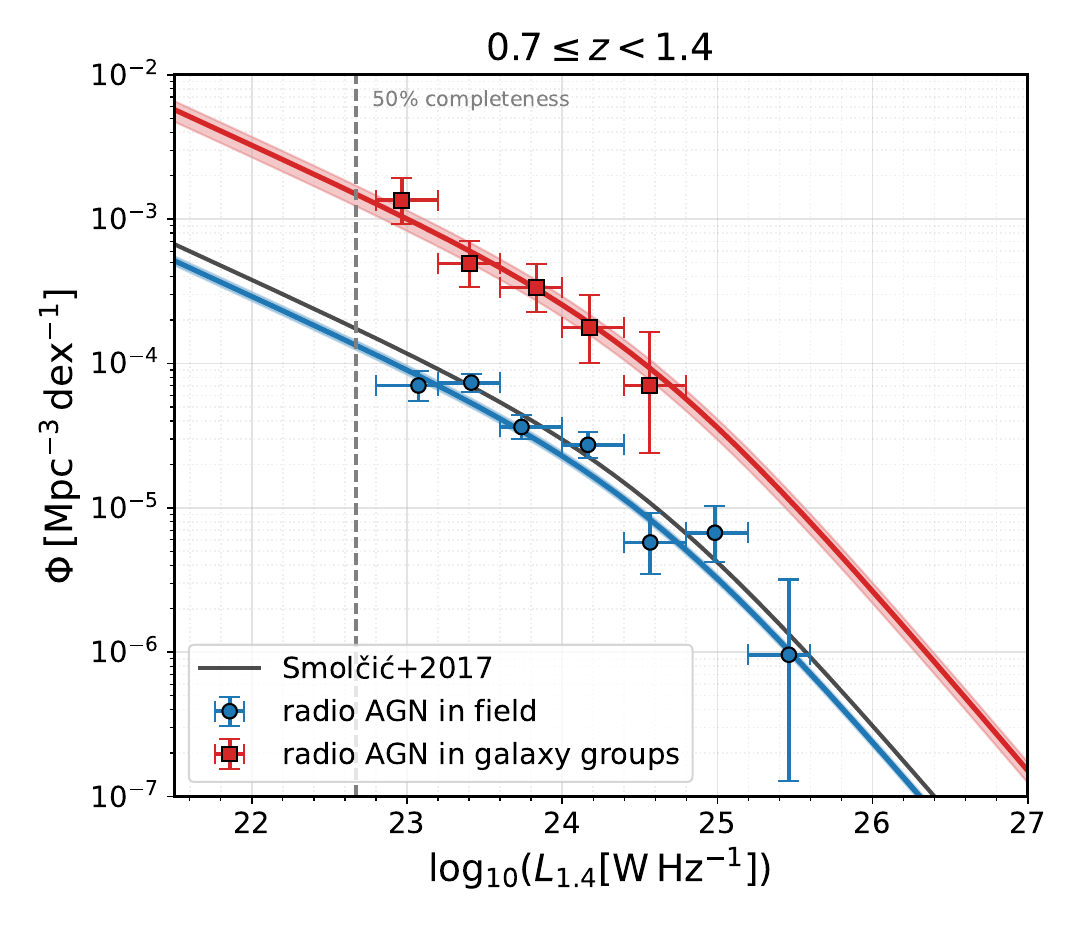} \\
        \includegraphics[width=0.5\linewidth]{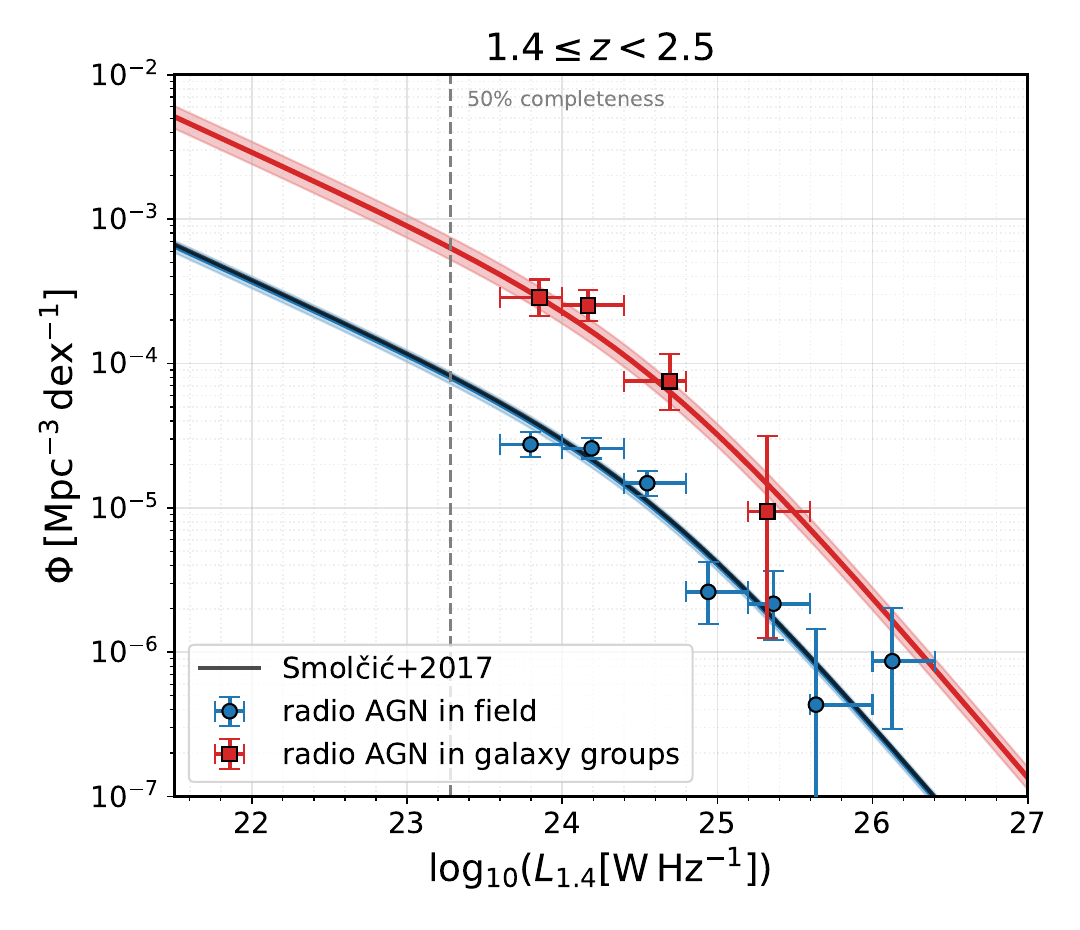}%
        \includegraphics[width=0.5\linewidth]{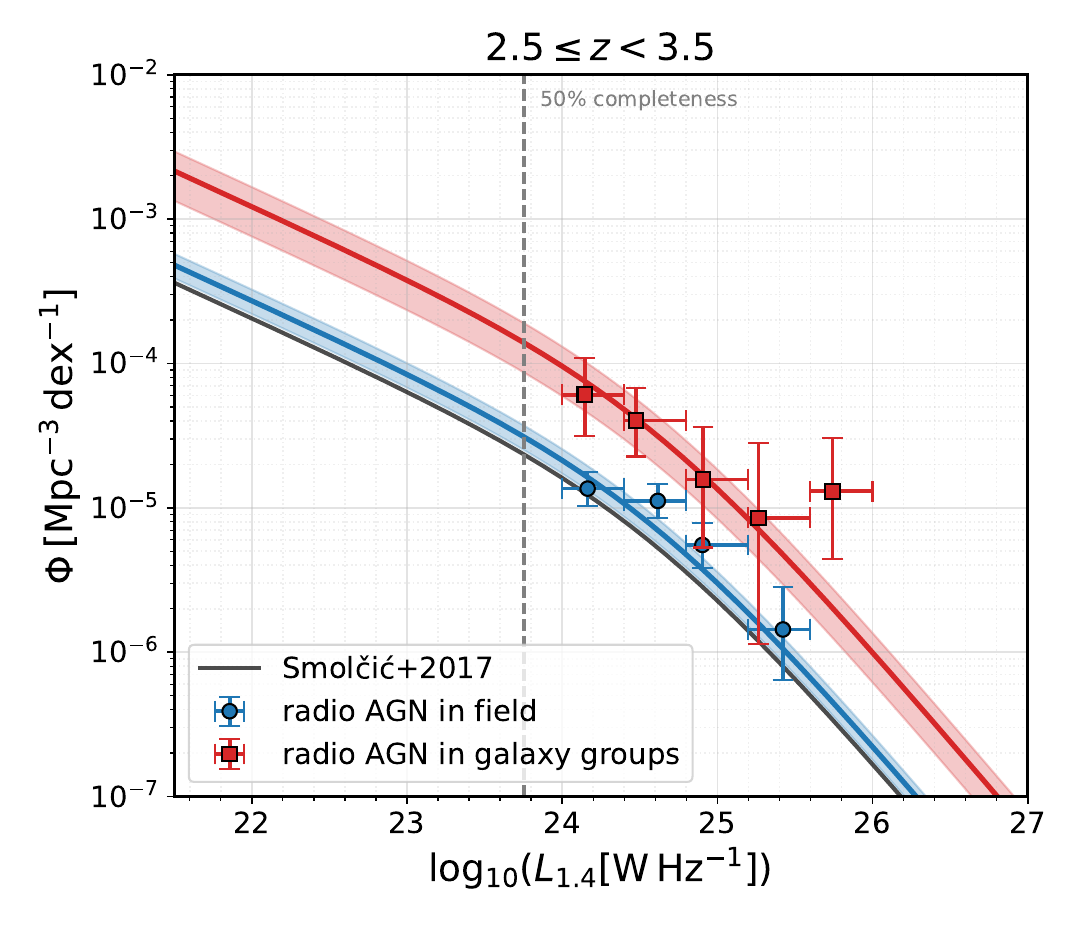}
    \end{minipage}%
    \hfill
    \begin{minipage}[c]{0.28\linewidth}
        \caption{The radio AGN luminosity function is shown for galaxies in the field (blue circles) and in galaxy groups (red squares) across different redshift bins (see Sect.~\ref{subsub:buildRLF}). Solid lines show the median posterior predictions from the pure density evolution (PDE) model described in Sect.~\ref{subsub:fitRLF} and reported in Table~\ref{tab:fitRLF}, while shaded regions denote the 16$^{\rm th}$–84$^{\rm th}$ percentile confidence intervals. The black curve and gray shaded band correspond to the best-fit PDE model and its uncertainty for the full sample of radio AGN in COSMOS from the 3 GHz VLA survey \citep{Smolcic2017_continuumdata,Smolcic2017_cosmicevo}. The vertical dashed line marks the luminosity corresponding to the flux density at the median redshift of the bin where the 50\% completeness level is reached. These plots show that the RLF of radio AGN in galaxy groups is systematically higher than in the field in the range $0.1\leq z<3.5$.}\label{fig:RLF}
    \end{minipage}
\end{figure*}
   
\section{Methods}\label{sec:methods}
The goal of this work is to investigate the demography of radio AGN activity in galaxies, with a focus on the influence of the environment. Therefore, we study the occurrence of radio AGN of varying rest-frame 1.4 GHz luminosity ($10^{22}~{\rm W~Hz^{-1}}\leq L_{\rm 1.4}\leq10^{26}~{\rm W~Hz^{-1}}$) as a function of redshift ($0.1\leq z \leq 3.5$), stellar mass ($10^{10}~{\rm M_{\odot}}\leq{\rm M}_{\ast}\leq2.5\times10^{11}~{\rm M_{\odot}}$),  and environment (galaxy groups or field). By performing a multi-parametric analysis, we wish to constrain the importance of each factor in triggering a radio AGN. In the following subsections, the bins of redshift, stellar mass, and radio luminosity were chosen to be large enough to contain a statistically significant number of galaxies. Redshift bins were chosen similarly to \citet{Delvecchio2022}, $0.1\leq z < 0.7$; $0.7\leq z < 1.4$; $1.4\leq z < 2.5$; and $2.5\leq z < 3.5$, in order to be roughly centered at representative redshifts of 0.5, 1, 2, and 3.

\subsection{Radio luminosity functions of radio AGN in the field and in galaxy groups out to z$\sim$3.5}\label{subsub:buildRLF}
In this subsection, we derive the radio AGN luminosity function (RLF) at 1.4~GHz over the redshift range $0.1 \leq z \leq 3.5$, separately for sources residing in galaxy groups and in the field.
The RLF of a population of $N$ radio AGN of radio luminosity $L$ at redshift $z$ represents the number density of radio-emitting sources per unit volume and luminosity interval, and can be written as:
\begin{equation}\label{eq:phiLz}
    \Phi (L,z) = \frac{1}{\Delta \log L}\sum_{i=1}^{N}\frac{1}{V_{max,i}}
\end{equation}
where $V_{max,i}$ is the maximum observable volume of the $i$-th radio AGN \citep{Schmidt1970}, and $\Delta \log L$ is the width of the luminosity bin. The rest-frame radio luminosity $L$ is usually computed at 1.4 GHz. Since the radio catalog employed in this work is at a reference frequency of 3~GHz, we scaled the luminosities to 1.4 GHz using the 1.4 - 3 GHz spectral index of each source (when available, for sources detected at both frequencies\footnote{About 62\% of sources in groups and 73\% of sources in the field; for these, we have $\langle\alpha_{groups}\rangle = 0.93\pm0.44$, and $\langle\alpha_{field}\rangle = 0.93\pm0.54$. Since the VLA-COSMOS 1.4 GHz survey \citep{Scoville2007} is shallower than the VLA-COSMOS 3 GHz survey (\citealt{Smolcic2017_continuumdata}; see also \citealt{Delhaize2017}), sources detected at both frequencies will have, on average, steeper spectra than the median $\alpha=0.7$ obtained including limits on the 1.4 GHz flux \citep{Smolcic2017_continuumdata}.} see \citealt{Scoville2007}), or by assuming the average spectral index $\alpha=0.7$ of VLA-COSMOS 3 GHz radio AGN from \citet{Smolcic2017_continuumdata} -- which was also adopted in \citet{Delvecchio2021} to build the IRRC.
Based on Eq.~\ref{eq:phiLz}, the calculation of the RLF separately for radio AGN in the field and in galaxy groups depends entirely on how the maximum observable volume is calculated. As detailed in the following, due to the intrinsically different spatial distributions in the Universe of galaxies in groups and in the field, we adopt two different approaches to model the RLF of radio AGN within the two environments.

\paragraph{V$_{max}$ for radio AGN in field --}
Following \citet{Novak2017}, we computed the maximum observable volume $V_{max,i}$ of the i-th source and applied corrections for radio AGN incompleteness (see also e.g., \citealt{Smolcic2017_cosmicevo,Delvecchio2022,Vardoulaki2025}). In particular, the maximum observable volume of a radio AGN that falls within the redshift bin $[z_{min},z_{max}]$ can be defined as:
\begin{equation}\label{eq:Vmaxfield}
    V_{max,i} = \sum_{z=z_{min}}^{z'} [V(z+\Delta z) - V(z)]C(z)
\end{equation}
where the sum starts at the beginning of the chosen redshift bin
and sums spherical shells of comoving volume $\Delta V = V(z+\Delta z) - V(z)$, computed in steps of $\Delta z = 0.01$, up to $z'$. The redshift $z'$ is defined as the minimum between the upper edge of the redshift bin, $z_{\rm max}$, and the maximum redshift at which a source of luminosity $L_i$ has an apparent flux density $S_{\rm 3\,GHz}(z)$ that remains above an adopted flux-density completeness limit. In this work, we use the completeness corrections, $C_{radio}$, from \citet{Smolcic2017_continuumdata} and we integrate volumes down to a completeness level of $C_{radio} = 50\%$. In Eq.~\ref{eq:Vmaxfield}, the factor $C(z)$ encapsulates the area of the survey and the radio incompleteness, such that:
\begin{equation}\label{eq:czfield}
    C(z) = \frac{A_{\rm C-Web}}{A_{\rm sky}}\times C_{radio}[S_{\rm 3\,GHz}(z)]\, ,
\end{equation}
where $A_{\rm C-Web} = 0.45$~deg$^{2}$ is the effective COSMOS-Web area and $A_{\rm sky} = 41253$~deg$^{2}$ is the area of the entire sky. 

\paragraph{V$_{max}$ for radio AGN in galaxy groups --} For radio AGN in galaxy groups, the maximum volume is different. Here, we are interested in the number density of radio sources within specific cosmic structures, that is, the probability of a given radio AGN of radio luminosity $L$ at redshift $z$ to be found within any galaxy group in the considered redshift bin. 
To do so, we considered that in this case the maximum observable volume of the $i$-th radio AGN is:
\begin{equation}\label{eq:Vmaxgroup}
    V_{max,i}\equiv \sum_{j=1}^{m} V^{\text{group}}_{j} \times C_{radio}[S_{\rm 3\,GHz}(z_{j})]\times G(z_j,\lambda^{\ast}_j)\, ,
\end{equation}
where $V^{\text{group}}_{j}$ is the volume of the $j$-th galaxy group within the redshift bin (see below). The sum over groups is performed for all groups for which the $i$-th radio AGN of luminosity $L_i$ could be observed while remaining above the same flux limit adopted for the field (i.e. corresponding to a completeness of 50\%). The factor $C_{radio}[S_{\rm 3\,GHz}(z_{j})]$ is the same as in Eq.~\ref{eq:czfield}. The factor $G(z_j,\lambda^{\ast}_j)$ encapsulates the corrections for purity and completeness of each group of redshift $z_{j}$ and intrinsic richness $\lambda^{\ast}_{j}$ (from \citealt{Toni2025}, as detailed in Sect.~\ref{subsec:cosmosgroup}): 
\begin{equation}\label{eq:grouppurcomp}
    G(z_{j},\lambda^{\ast}_{j}) = \frac{\text{Purity}\,(z_{j},\lambda^{\ast}_{j})}{\text{Completeness}\,(z_{j},\lambda^{\ast}_{j})}
\end{equation}
The group volume $V^{\text{group}}_{j}$ is defined as the comoving volume
enclosed within the spatial extent of the group:
\begin{equation}\label{eq:volumegroups}
    V^{\text{group}}_{j} =\frac{4\pi}{3}\times (1+z)^{3}\times R_{95\%}^{3}\, 
\end{equation}
where $R_{95\%}$ corresponds to twice the root mean square of the group-centric galaxy distribution (i.e., a $\sim$2$\sigma$, or 95\% extent). In computing $R_{95\%}$, the distances from the group center of all galaxies associated with the group (of any stellar mass or with any \texttt{ASSOC\_PROB}) are weighted by their membership probability (\texttt{ASSOC\_PROB}), following the approach described in \citet{Toni2026} (see their Appendix~B). Typical values are $R_{95\%} \sim 0.5$--$1$~Mpc. $V_{{\rm max},i}$ can thus be interpreted as the sum of the effective volumes of all groups that could host and allow the detection of a radio AGN with luminosity $L_i$, accounting for both radio completeness and the systematics of the group catalog\footnote{Formally, the total volume occupied by galaxy groups should be subtracted from the field volume within each redshift bin. However, implementing this correction is not straightforward because $V_{{\rm max},i}$ is computed individually for each field radio AGN. In practice, the correction is negligible, as the total group volume represents only $0.09\%$ ($0.1\leq z<0.7$), $0.13\%$ ($0.7\leq z<1.4$), $0.19\%$ ($1.4\leq z<2.5$), and $0.34\%$ ($2.5\leq z<3.7$) of the corresponding total survey volume.}. This definition intentionally adopts a conservative estimate of the group volume. Since $R_{95\%}$ encloses the probabilistic spatial extent of a galaxy group based on all its associated galaxies, irrespective of their $M_{\ast}$ or membership probability, it provides an approximate upper limit to the physical volume occupied by each group.

\par We used the catalogs \say{${\rm GRP_{rAGN}}$} and \say{${\rm FLD_{rAGN}}$} to build the RLFs for radio AGN in galaxy groups and for radio AGN in the field, respectively. The resulting RLFs are plotted in Fig.~\ref{fig:RLF}. We report
RLFs using the median $L_{1.4}$ in each radio luminosity bin, with the error
bar on the x-axis showing the width of the luminosity bin, and the error bar on the y-axis corresponding to Gehrels uncertainties \citep{Gehrels1986}. In Fig.~\ref{fig:RLF} we also overlay the best-fit RLF of radio AGN in the VLA-COSMOS 3 GHz survey modeled with a pure density evolution from \citet{Smolcic2017_cosmicevo}. We note here that, since this reference RLF is derived from the global radio AGN population in VLA-COSMOS 3~GHz -- numerically dominated by field sources\footnote{Based on Fig.~\ref{fig:venn}, the radio AGN in the COSMOS-Web area that reside in the field are $\sim$76\% of the total.} -- we expect it to more closely resemble the RLF of field radio AGN than that of group radio AGN. This expectation is supported by Fig.~\ref{fig:RLF}, where the global RLF is in better agreement with the field population.

\section{Analysis and Results}\label{sec:analysis}
\subsection{Fitting the radio luminosity function}\label{subsub:fitRLF}
In order to evaluate differences between the RLF of radio AGN in groups and field, as well as to understand how the RLFs evolve over cosmic time, we performed a fit of the observed RLFs shown in Fig.~\ref{fig:RLF}. In general, the redshift evolution of radio AGN populations as a function of luminosity $L$ and redshift $z$ is usually expressed as a joint density and luminosity evolution of its local luminosity function (e.g., \citealt{Smolcic2017_cosmicevo,Delvecchio2022,Kondapally2022,vanderVlugt2022,Wang2024,Vardoulaki2025}). Fitting a model that includes both luminosity and density evolution parameters usually requires very large statistics; for example, \citet{Smolcic2017_cosmicevo} showed that there is a strong degeneracy between the luminosity and density evolution parameters in fitting the RLF of the global radio AGN population in COSMOS. Here, given the smaller area of COSMOS-Web and our separation of radio AGN between field and groups, we decided to assume a pure-density evolution (PDE) to fit the observed RLFs. While the \textit{true} cosmic evolution of radio AGN in groups and field might be different, we are mostly interested in the difference between the two populations assuming the \textit{same} cosmic evolutionary model. This approach also enables direct comparison with previous studies and facilitates the study of how the relative radio AGN space densities in groups and in the field evolve with cosmic time. Therefore, we assume the following pure density evolution:
\begin{equation}\label{eq:PDE}
    \Phi(L,z) = (1+z)^{\alpha_{D}}\times \Phi_{0}(L)\,,
\end{equation}
where $\alpha_{D}$ is the characteristic density evolution parameter and $\Phi_{0}(L)$ is the RLF in the local Universe. This form is usually fitted individually to the observed RLF in each redshift bin (e.g., \citealt{Smolcic2017_cosmicevo,Wang2024}); additionally, a continuous model in the form:
\begin{equation}\label{eq:PDEcont}
    \Phi(L,z) = (1+z)^{(\alpha_{D} + z\cdot\beta_{D})}\times \Phi_{0}(L)\,,
\end{equation}
can also be used to fit the data simultaneously in all redshift bins with a redshift dependent term, $\alpha_{D} + z\cdot\beta_{D}$, that substitutes the constant $\alpha_{D}$ of Eq.~\ref{eq:PDE}.
The local RLF in the above equations, $\Phi_{0}(L)$, can be expressed analytically as \citep{MauchSadler2007}:
\begin{equation}\label{eq:localRLF}
    \Phi_{0}(L) = \frac{\Phi^{\ast}}{(L/L^{\ast})^{\delta_{1}} + (L/L^{\ast})^{\delta_{2}}}\, ,
\end{equation}
where $\Phi^{\ast} = \frac{1}{0.4}10^{-5.5}$~Mpc$^{-3}$~dex$^{-1}$ is the normalization, $L^{\ast} = 10^{24.59}$~W~Hz$^{-1}$ is the position of the knee, and $\delta_{1} = -1.27$ and $\delta_{2} = -0.49$ are the bright and faint end slopes, respectively.

When fitting the RLF, we employed a Markov Chain Monte Carlo (MCMC) approach using the Python package \texttt{emcee}, adopting uniform priors on all parameters and allowing them to vary over unconstrained ranges. We also excluded the lowest luminosity bin whenever its median $L_{1.4}$ was consistent with the 50\% radio completeness limit of the redshift bin (these bins are not plotted in Fig.~\ref{fig:RLF}). In typical RLF studies of the global radio AGN population, it is common to assume the local RLF as in Eq.~\ref{eq:localRLF}, from \citet{MauchSadler2007}, and then fit for the cosmic evolution (either in each $z$ bin through Eq.~\ref{eq:PDE}, or with a global $z$ model as in Eq.~\ref{eq:PDEcont}). Here, instead, we may have to reconsider the local RLF, which may vary among different environments. 

For both field and group radio AGN, we initially fitted the observed RLFs with the global redshift evolution of Eq.~\ref{eq:PDEcont} and Eq.~\ref{eq:localRLF}, allowing all parameters to vary freely ($\alpha_{D}$, $\beta_{D}$, $\Phi^{\ast}$, $L^{\ast}$, $\delta_{1}$, and $\delta_{2}$). However, the large number of free parameters and the strong degeneracies among them prevented accurate constraints. In practice, this resulted either in poor reduced $\chi^2$ values or in parameter uncertainties large enough to remain consistent with the local RLF parameters reported by \citet{MauchSadler2007}. To mitigate these issues, we adopted an iterative fitting strategy. Starting from the fully free-parameter model, we progressively fixed the parameters describing the local RLF in the following order: $\delta_{1}$, $\delta_{2}$, $L^{\ast}$, and $\Phi^{\ast}$. At each step, we evaluated whether the best-fit parameters remained consistent, within uncertainties, with the local RLF values from \citet{MauchSadler2007}, and whether the reduced $\chi^2$ decreased. We found that this procedure led to stable and improved fits up to the point where $\Phi^{\ast}$ was still allowed to vary. Consequently, in our final fits for both group and field radio AGN, we let $\Phi^{\ast}$, $\alpha_{D}$, and $\beta_{D}$ free to vary, and we fixed the remaining ones ($\delta_1, \,\delta_2,\,L^{\ast}$). The resulting best-fit models are shown in Fig.~\ref{fig:RLF}, and the corresponding parameters and reduced $\chi^2$ values are reported in Table~\ref{tab:fitRLF}. Corner plots with the posterior distributions of the fitted parameters are shown in Fig.~\ref{fig:cornerRLF}.

Additionally, we tested the fit of the RLF in each redshift bin, with the PDE being described as in Eq.~\ref{eq:PDE}. In previous similar studies (e.g., \citealt{Smolcic2017_cosmicevo,Wang2024}), and as noted above for the global redshift fit, the local RLF of \citet{MauchSadler2007} is assumed and only $\alpha_{D}$ is left free to vary. Here, as noted above, we cannot fix the local RLF (especially its normalization, $\Phi^{\ast}$) to the values of \citet{MauchSadler2007}. At the same time, the fit of each individual bin implies a smaller statistics relative to the global fit that, when combined with the strong degeneracy between $\Phi^{\ast}$ and $\alpha_{D}$, prevents us from obtaining stable best-fit results. Therefore, we froze $\Phi^{\ast}$ to the best-fit obtained from the global fit (see Table~\ref{tab:fitRLF}), and left $\alpha_{D}$ free to vary. The results of the individual-$z$ fits are reported in Tab~\ref{tab:fitRLF-zbins}.

\begin{figure}[ht!]
    \centering
    \includegraphics[width=0.8\linewidth]{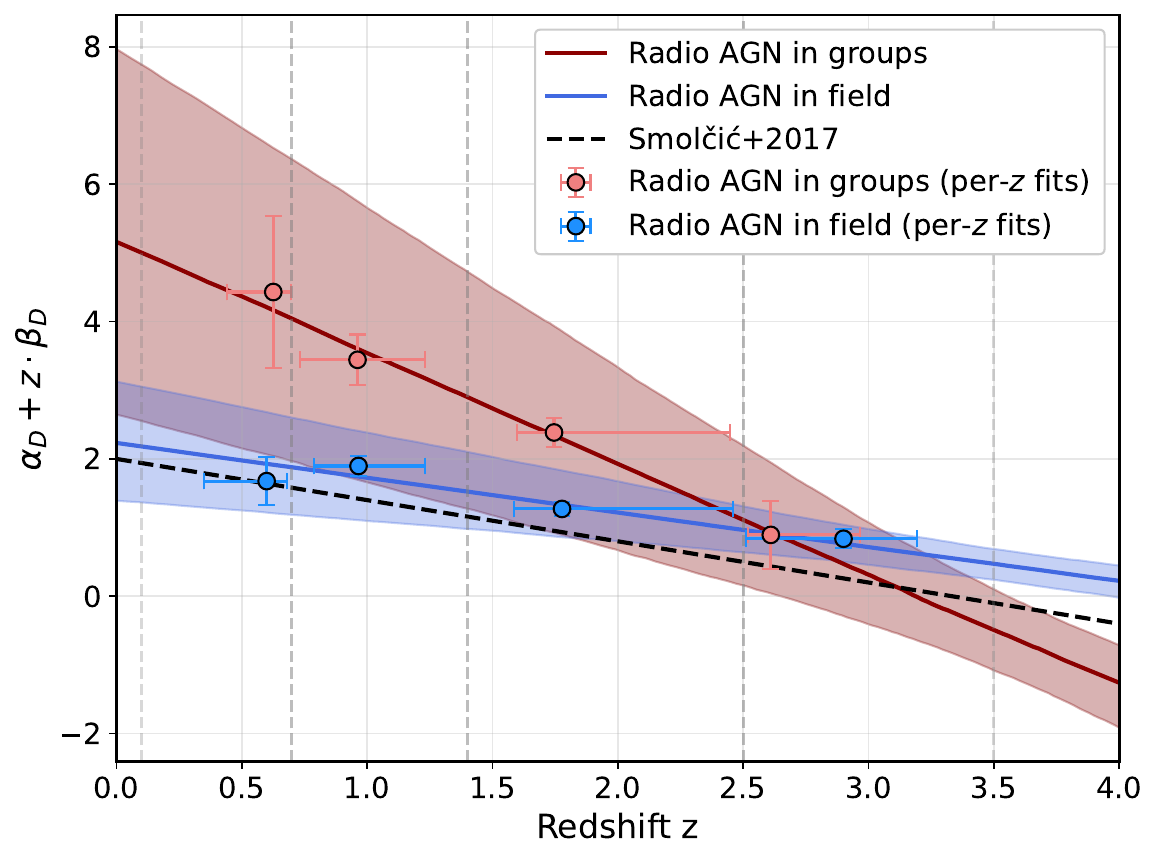}\vspace{-0.2cm}
    \caption{Redshift evolution of the best-fit PDE parameters $\alpha_{D}$ and $\beta_{D}$, as defined in Eq.~\ref{eq:PDEcont}. We plot $\alpha_{D} + z\cdot\beta_{D}$ for radio AGN in groups (red), radio AGN in field (blue) and radio AGN in COSMOS2020 from \citet{Smolcic2017_cosmicevo} (dashed black). We also overlay with red and blue circles the best-fit per-z $\alpha_{D}$, as defined in Eq.~\ref{eq:PDE}. 
    }
    \label{fig:RLFevo}\vspace{-0.2cm}
\end{figure}

In Figure \ref{fig:RLFevo}, we plot the best-fit $\alpha_{D} + z\cdot\beta_{D}$ for radio AGN in groups and field as a function of redshift (see also Table~\ref{tab:fitRLF}). We also overlay, for comparison, the best-fit $\alpha_{D} + z\cdot\beta_{D}$ PDE from \citet{Smolcic2017_cosmicevo}. Additionally, we plot the results of the fit to each individual redshift bin as in Eq.~\ref{eq:PDE} (as reported in Table~\ref{tab:fitRLF-zbins}). We observe a general agreement of the best-fit cosmic evolution for each population derived from an individual redshift bin (Eq.~\ref{eq:PDE} and Table~\ref{tab:fitRLF-zbins}) and from a global (Eq.~\ref{eq:PDE} and Table~\ref{tab:fitRLF}) fitting procedure. Within uncertainties, the redshift evolution has a similar shape between radio AGN in field, in galaxy groups, and the overall population, although the population of radio AGN in galaxy groups might have a steeper evolution (i.e., a faster decrease of density with redshift). We will return to this point in Sect.~\ref{subsec:disc-overdensities}, where we will interpret these results together with the results from the analysis of the radio AGN fraction.

\begin{figure*}[htp!]
    \centering

    \includegraphics[width=0.95\linewidth]{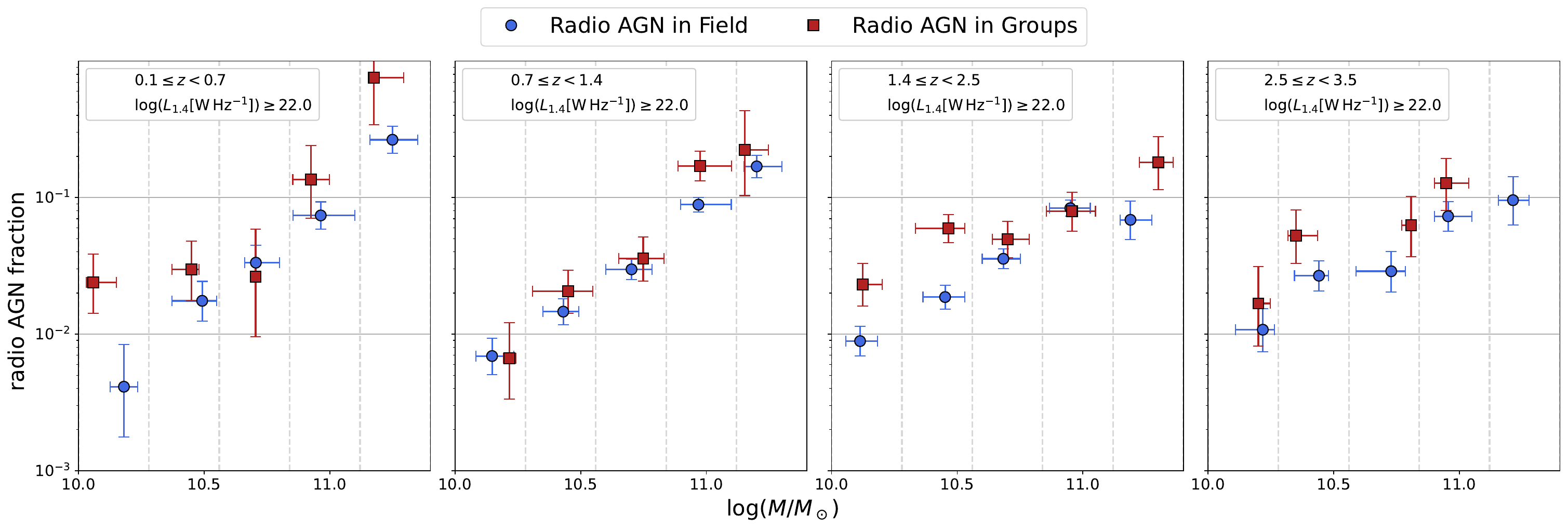}

    \caption{Radio AGN fractions as a function of stellar mass for field (blue circles) and group (red squares) environments. Radio AGN fractions are calculated for $L_{1.4} \geq 10^{22}$~${\rm W\,Hz^{-1}}$ in four redshift bins between $0.1 \leq z \leq 3.5$. Points are plotted at the median stellar mass of each bin. Vertical dashed lines mark the stellar mass binning used in the analysis. In addition to the increase with stellar mass in both environments, the radio AGN fractions in galaxy groups are equal to or higher than in the field across all four redshift bins.}
    \label{fig:fractions}
\end{figure*}

\subsection{Radio AGN fractions as a function of $M_{\ast}$, $z$, $L_{1.4}$}\label{subsec:fractions}
In this subsection, we use a multi-parametric approach to simultaneously investigate the dependence of the radio AGN fraction on the environment, on redshift ($0.1\leq z \leq 3.5$), on stellar mass ($10.0\leq\log({\rm M}_{\ast}{\rm [M_{\odot}]})\leq11.4$), and on  radio luminosity $22\leq\log({\rm L_{1.4}}{\rm [W~Hz^{-1}]})\leq26$. To do so, we first calculate the probability of a galaxy at redshift $z$, of stellar mass $M_{\ast}$ in a given environment to host a radio AGN of luminosity $L_{1.4}$. This is simply the radio AGN fraction, defined as:

\begin{equation}\label{eq:fAGNgroups}
    f_{\rm AGN}\big\rvert_{\rm env}\,(M_{\ast},z,L_{1.4}) = \frac{N_{\rm radio\, AGN}\in {\rm env} \,(M_{\ast},z,L_{1.4})}{N_{\rm gal,tot}\in {\rm env} \,(M_{\ast},z)}
\end{equation}

where $N_{\rm radio\,AGN}$ is computed as the sum over radio AGN sources, weighted by the inverse of the radio completeness, i.e. each source contributes a factor $1/C_{\rm radio}[S_{\rm 3\,GHz}(z)]$ following \citet{Smolcic2017_continuumdata}, while $N_{\rm gal,tot}$ corresponds to the total number of galaxies in the given bin and environment, including both radio AGN and non radio AGN galaxies. Completeness corrections were applied down to 50\% completeness in flux density; stellar mass bins in which the median radio luminosity is consistent (within uncertainties) with this limit are conservatively excluded. In practice, the fractions above can be expressed as the following combination of the catalogs listed in Sect.~\ref{subsec:createcat}:
\begin{equation}\label{eq:fAGNgroups-cat}
    \frac{\rm  GRP_{rAGN}}{\rm GRP_{rAGN} + GRP_{GAL}}\bigg\rvert_{M_{\ast},z,L_{1.4}} \rightarrow  f_{\rm AGN}\big\rvert_{\rm groups}\,(M_{\ast},z,L_{1.4})
\end{equation}
\begin{equation}\label{eq:fAGNfield-cat}
    \frac{\rm FLD_{rAGN}}{\rm FLD_{rAGN} + FLD_{GAL}}\bigg\rvert_{M_{\ast},z,L_{1.4}} \rightarrow  f_{\rm AGN}\big\rvert_{\rm field}\,(M_{\ast},z,L_{1.4})
\end{equation}

\noindent The fractions plotted in Fig.~\ref{fig:fractions} and Fig.~\ref{fig:fractionsall} were computed using the same redshift bins as in the study of the RLF, and by splitting our radio AGN into five equally spaced stellar mass bins in the range $10.0\leq\log({\rm M}_{\ast} [\rm M_{\odot}])\leq11.4$. We verified that the fraction of star-forming galaxies contributing to the radio AGN population does not differ significantly between group and field environments, within the current statistical uncertainties This indicates that any excess or deficit in the radio AGN fractions observed in groups relative to the field is not driven by differences in the star-forming galaxy population. In Fig.~\ref{fig:fractions}, we plot radio AGN fractions over the full range of radio power ($L_{\rm 1.4} \geq 10^{22}~{\rm W~Hz^{-1}}$). In Fig.~\ref{fig:fractionsall}, we further separate the data into four luminosity bins of 1~dex width between $10^{22}~{\rm W~Hz^{-1}}$ and $10^{26}~{\rm W~Hz^{-1}}$.
Points are plotted at the median ${\rm M}_{\ast}$, and the uncertainties on the x-axis represent the 16$^{\rm th}$- and 84$^{\rm th}$-percentile of the stellar mass distribution in each bin. Uncertainties on the radio AGN fractions are computed following \citet{Gehrels1986}.

A clear trend that is readily visible from the plots of Fig.~\ref{fig:fractions} and Fig.~\ref{fig:fractionsall} is that the radio AGN fraction increases with stellar mass at all redshifts and radio luminosities. In Fig.~\ref{fig:fractions}, we see that at the highest stellar masses ($M_{\ast} \sim 2 \times 10^{11}\,{\rm M_{\odot}}$) in the lowest redshift bin the fractions approach $\sim 50-100\%$, indicating that nearly all local massive galaxies host low-power radio AGN. This is consistent with previous findings (e.g. \citealt{Best2005,Sabater2019,Wang2024}). In addition, we find that the radio AGN fraction in galaxy groups is on average a factor of 2 higher than in the field across all redshift bins explored here, both when separating in bins of $L_{1.4}$ and when considering the full $L_{1.4}\geq10^{22}$~W~Hz$^{-1}$ luminosity range. While this environmental dependence of the radio AGN fraction has been previously observed from the local Universe (e.g. \citealt{Sabater2013}) to $z\leq1.5$ \citep{Mo2020,Hashiguchi2023}, our results indicate for the first time that it persists out to $z \sim 3.5$. This suggests that, at fixed $M_{\ast}$ and $L_{1.4}$, dense environments enhance the triggering of radio AGN over a wide range of cosmic time. We provide a parametric description of these trends in Sect.~\ref{subsec:fitfractions}, and we further discuss the implications of our results in Sect.~\ref{subsec:disc-overdensities}.

\begin{figure}[htp!]
    \centering
    \includegraphics[width=0.78\linewidth]{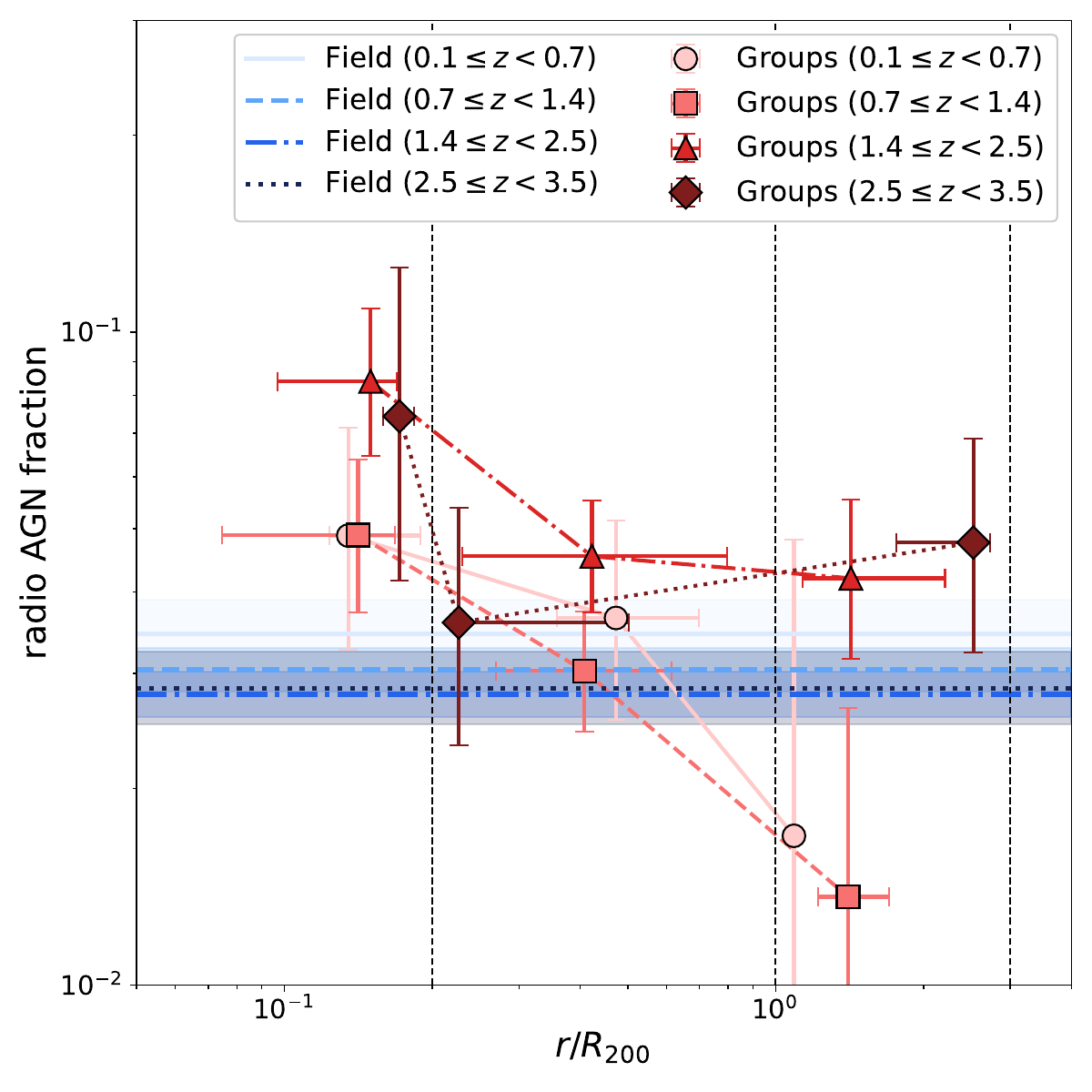}
    \caption{Radio AGN fractions (computed with Eq.~\ref{eq:fAGNgroups-cat}) as a function of projected group-centric distance $r/R_{200}$ (Eq.~\ref{eq:R200}), (darker red corresponds to increasing $z$). Red points represent the radio AGN fraction in groups, while horizontal lines (and shaded region) indicate the radio AGN fraction in the field (Eq.~\ref{eq:fAGNfield-cat}). Vertical dashed lines mark the boundaries of the radial bins ($0.2R_{200}$, $R_{200}$, $3R_{200}$). Points are plotted at the median $r/R_{200}$, with uncertainties on the x-axis representing the 16$^{\rm th}$- and 84$^{\rm th}$-percentile. 
    }
    \label{fig:fractions-radius}
\end{figure}

\subsection{Radial distribution of radio AGN fractions in galaxy groups}\label{subsec:AGNfraction-vsradius}
In this subsection, we study how the fraction of group galaxies that are radio AGN varies with the distance of the galaxies from the group center. Most previous studies have been limited to overdensities at $z \leq 1.5$; here we extend the analysis to $z \sim 3.5$.

We computed the fraction of group galaxies hosting a radio AGN across the full stellar mass and radio luminosity range probed by our sample, applying corrections for radio incompleteness. The analysis is performed in bins of redshift (defined in Sect.~\ref{sec:methods}) and projected radial distance from the group center, expressed in units of $r/R_{200}$. The projected distance $r$ is measured relative to the group centers provided in the catalog of \citet{Toni2025}.  To define a representative $R_{200}$ in each redshift bin, we compute the median halo mass $M_{200}^{\rm med}$ and median redshift $z^{\rm med}$ of the groups, and derive:
\begin{equation}\label{eq:R200}
    R_{200} = \left(\frac{3 M_{200}^{\rm med}}{4  \pi \times 200  \rho_{\rm crit}(z^{\rm med})}\right)^{1/3}  
    = \left(\frac{M_{200}^{\rm med}\times  G}{100 \times H(z^{\rm med})^{2} }\right)^{1/3}\,,
\end{equation}
where $G$ is the gravitational constant. We adopted radial bins of $r/R_{200} = [0, 0.2],\, [0.2, 1],\, [1, 3]$ to facilitate comparison with the literature (\citealt{Best2007,Smolcic2011}). We verified that splitting the sample into stellar mass bins yields the same trends discussed below, although with larger uncertainties due to smaller statistics. We also verified that the fraction of SFG does not change significantly between the radial bins. The fractions are plotted in Fig.~\ref{fig:fractions-radius}, where we also indicate the average radio AGN fraction measured in the field.

At $z \lesssim 1.4$, we observe an increase in the radio AGN fraction toward the group centers, with an enhancement within $r \lesssim 0.2\,R_{200}$ of up to a factor of $\sim 4$ with respect to $r \gtrsim R_{200}$. In this redshift range, the radio AGN fraction in groups is lower than in the field at $r \gtrsim R_{200}$, consistent with the field at $0.2\,R_{200} \lesssim r \lesssim R_{200}$, and higher in the central regions. These results align with existing studies of radio AGN occupation in groups (\citealt{Best2007} for $z\leq0.1$; \citealt{deVos2024} for $z\leq0.2$; \citealt{Smolcic2011} for $z\leq1.3$ X-ray selected groups and clusters; \citealt{Hashiguchi2023} for $z\leq1.3$ optically-selected groups and clusters). At higher redshifts ($z \gtrsim 1.4$), the trend appears to change. While the radio AGN fraction remains highest at $r \leq 0.2\,R_{200}$, there seems to be a flattening out to $r \sim 3\,R_{200}$. This suggests that, at earlier cosmic times, radio AGN triggering may also be efficient in galaxy groups outskirts (e.g., \citealt{Bigwood2024}). We further discuss these trends in Sect.~\ref{subsec:disc-overdensities}.

\begin{figure}[ht!]
    \centering
    \includegraphics[width=0.8\linewidth]{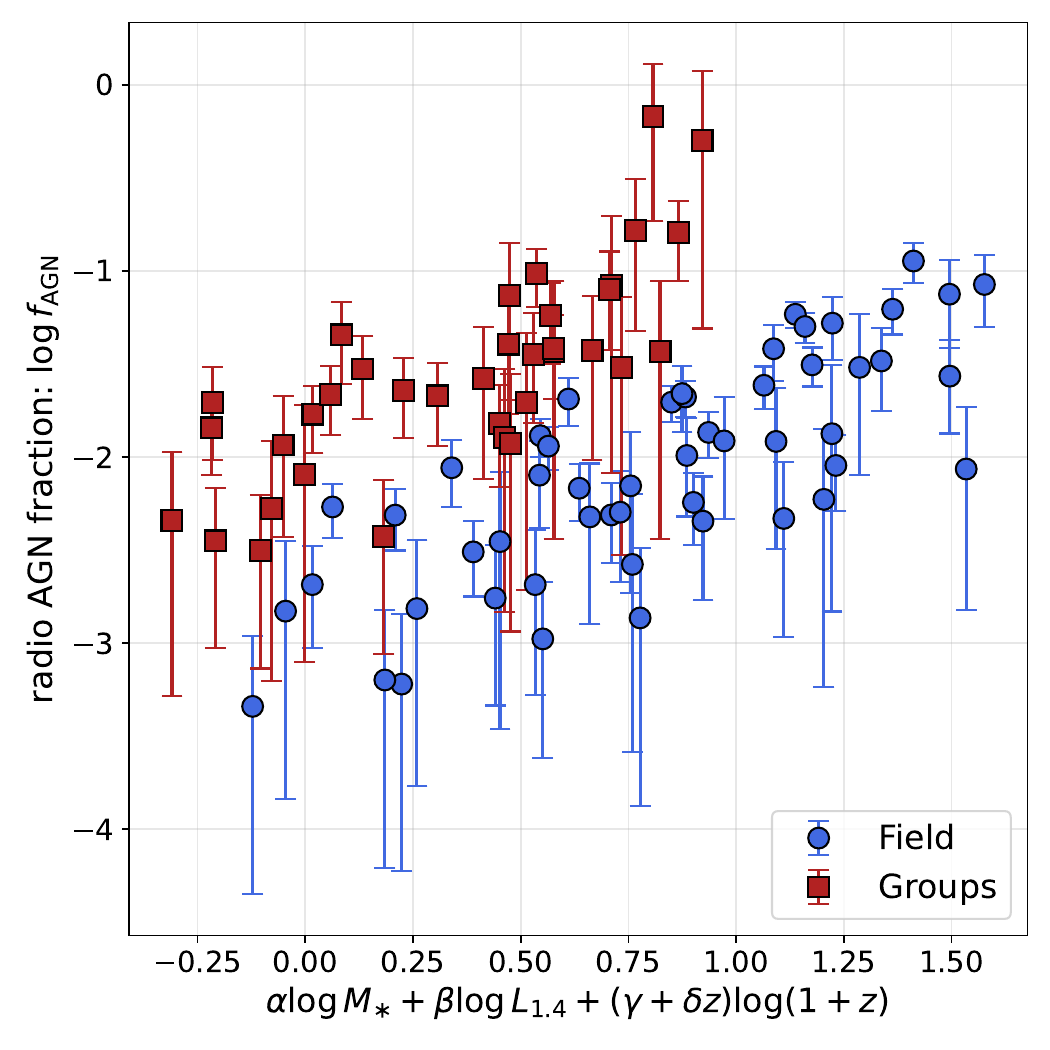}
    \caption{Radio AGN fractions from Fig.~\ref{fig:fractionsall} plotted against the projected x-coordinate $x = \alpha\log\left(\frac{M_{\ast}}{10^{10}\,\rm M_{\odot}}\right) + \beta\log\left(\frac{L_{1.4}}{10^{22}\,\rm W~Hz^{-1}}\right) + (\gamma + \delta z)\log(1+z)$, where $\alpha$, $\beta$, $\gamma$, $\delta$ are the best-fit values reported in Table~\ref{tab:bestfit-fractions}. Using a plane to describe the individual dependencies on $M_{\ast}$, $L_{1.4}$, and $z$ highlights the remaining offset between radio AGN fractions in galaxy groups and in the field.}
    \label{fig:bestfit-allfractions}
\end{figure}

\subsection{A new parametric form to fit the fractions of radio AGN}\label{subsec:fitfractions}
In order to provide a quantitative estimate of the trends of radio AGN fraction versus stellar mass, radio luminosity, redshift, and environment, we proceeded to fit the AGN fractions displayed in Fig.~\ref{fig:fractionsall} with a parametric form. Specifically, based on the qualitative trends discussed in Sect.~\ref{subsec:fractions} and on the plots of Fig.~\ref{fig:fractionsall}, and informed by previous studies in the literature (e.g., \citealt{Wang2024}), we assume that the radio AGN fraction can be described as the combination of power-laws in the log-log space for $M_{\ast}$ and $L_{1.4}$, and a redshift-dependent evolutionary term. This last term is added due to the non-monotonic changes in the radio AGN fractions with redshift, visible in Fig.~\ref{fig:fractions} and \ref{fig:fractionsall}. Specifically, we model the radio AGN fraction as a function of $M_{\ast}$, $L_{1.4}$, and $z$ for a fixed environment as:
\begin{equation}\label{eq:agnfractionfit}
\begin{split}
    & f_{\rm AGN} (M_{\ast},L_{1.4},z\,\rvert \,{\rm env}) = \\
    & 10^{\rm const.}\left(\frac{M_{\ast}}{10^{10}\,\rm M_{\odot}}\right)^{\alpha} 
     \left(\frac{L_{1.4}}{10^{22}\,\rm W~Hz^{-1}}\right)^{\beta} 
     (1+z)^{(\gamma + \delta z)}\,,
\end{split}
\end{equation}
where $\alpha$, $\beta$, $\gamma$, $\delta$, and ${\rm const.}$ are the free parameters. 
The model is fit to the observed radio AGN fractions in the log-log space, assuming Gaussian uncertainties on the radio AGN fractions. Rather than using the binned quantities, each point from Fig.~\ref{fig:fractionsall} is evaluated at its median $M_{\ast}$, $L_{1.4}$ and $z$. We adopt uniform priors on all parameters, allowing them to vary over unconstrained ranges, and then sample the posterior distribution using the MCMC sampler implemented in the \texttt{emcee} package. The posterior distributions are used to derive median best-fit parameters and the associated confidence intervals. The fit is performed separately for radio AGN in galaxy groups and for radio AGN in field. We report in Table~\ref{tab:bestfit-fractions} the best-fit parameters and statistics for this fit, and we report in Fig.~\ref{fig:corner-fractions} the corner plots for the best-fit parameters. The best-fit parametric form is overlaid in each subplot of Fig.~\ref{fig:fractionsall}, separately for radio AGN in galaxy groups and in the field. Additionally, in Fig.~\ref{fig:bestfit-allfractions}, we plot a \say{plane} representation of the dependence of $\log f_{\rm AGN}$ on $\log M_{\ast}$, $\log L_{1.4}$, and $z$. Specifically, we define the projected x-axis as $x = \alpha\log\left(\frac{M_{\ast}}{10^{10}\,\rm M_{\odot}}\right) + \beta\log\left(\frac{L_{1.4}}{10^{22}\,\rm W~Hz^{-1}}\right) + (\gamma + \delta z)\log(1+z)$, where the coefficients are taken from the best-fit of the radio AGN fractions (Table~\ref{tab:bestfit-fractions}), and then we plot the $\log f_{\rm AGN}$ versus this combined variable for field and group environments separately. Our best-fit is expressed as:
\begin{equation}\label{eq:agnfractionfitfield}
\begin{split}
    & f_{\rm AGN} (M_{\ast},L_{1.4},z\,\rvert \,{\rm field}) = \\
    & 0.0026\left(\frac{M_{\ast}}{10^{10}\,\rm M_{\odot}}\right)^{1.19}  \left(\frac{L_{1.4}}{10^{22}\,\rm W~Hz^{-1}}\right)^{-0.26}  (1+z)^{(1.31 -0.13 z)} \\
\end{split}
\end{equation}
\vspace{-0.5cm}
\begin{equation}\label{eq:agnfractionfitgroups}
\begin{split}
    & f_{\rm AGN} (M_{\ast},L_{1.4},z\,\rvert \,{\rm group}) = \\
    &  0.017 \left(\frac{M_{\ast}}{10^{10}\,\rm M_{\odot}}\right)^{0.95}
    \left(\frac{L_{1.4}}{10^{22}\,\rm W~Hz^{-1}}\right)^{-0.05}
    (1+z)^{(-1.52 + 0.46 z)} \\
\end{split}
\end{equation}
We observe that our best fit parametric fit correctly reproduces the trends of $f_{\rm AGN}$ increasing with stellar mass both in galaxy groups ($\alpha \simeq 0.95$) and in the field ($\alpha \simeq 1.19$). These slopes are smaller than those reported by \citet{Best2005} for radio AGN in the local Universe ($\alpha\sim2.5$ at $z\lesssim0.3$), but are consistent with more recent studies over broader $z$ ranges, such as \citet{Wang2017} ($\alpha\sim1.1$ over $0\leq z\leq4$) and \citet{Williams2015} ($\alpha\sim1.0$ at $1.5\leq z\leq2$). At fixed stellar mass and redshift, the dependence on radio luminosity is slightly different (but consistent at $1.4\sigma$) between the two environments. In the field, the radio AGN fraction seems to decrease with increasing $L_{1.4}$ ($\beta \simeq -0.26$), while in galaxy groups the dependence is consistent with being flat ($\beta \simeq -0.05$). If such a difference were real, it would suggest that, in dense environments, the probability of hosting a radio AGN may be less sensitive to radio power at fixed stellar mass -- similarly to the results of \citet{Best2007} for cluster-central radio AGN in the local Universe. 
\\\indent We caution that our model does not fully explain the data in all stellar-mass, radio-luminosity, or redshift bins. In particular, the model tends to over-predict the occurrence of: low luminosity ($22\leq \log (L_{1.4}{\rm [W~Hz^{-1}]})\leq23$) radio AGN in galaxy groups and in the field in the range $0.7< z<1.4$; high luminosity ($25\leq \log (L_{1.4}{\rm [W~Hz^{-1}]})\leq26$) radio AGN in the field in the range $1.4< z<2.5$; low luminosity ($23\leq \log (L_{1.4}{\rm [W~Hz^{-1}]})\leq24$) radio AGN in the field in the redshift range $2.5< z<3.5$ (see Fig.~\ref{fig:fractionsall}). 
In this context, we note that the adopted parametric form assumes separable power-law dependencies on stellar mass, radio luminosity, and redshift. However, we cannot exclude the presence of higher-order dependence between these quantities, such as a redshift dependence of $\alpha,\,\beta$, or a luminosity or mass dependence of $\alpha,\,\beta,\,\gamma\,,\delta$ (see also \citealt{Wang2024} for a similar discussion). 
Detailed testing of more complex models would require larger samples than available. We do present an initial study of higher-order dependencies between $f_{\rm AGN}$, $M_{\ast}$, $L_{1,4}$, and $z$ in Appendix \ref{app:fitfractions-complex}, but, given the uncertain and less motivated nature of more complex models, in the discussion we proceed with the simpler parametrization explored here.

\begin{figure*}[ht!]
        \centering
        \includegraphics[width=0.47\linewidth]{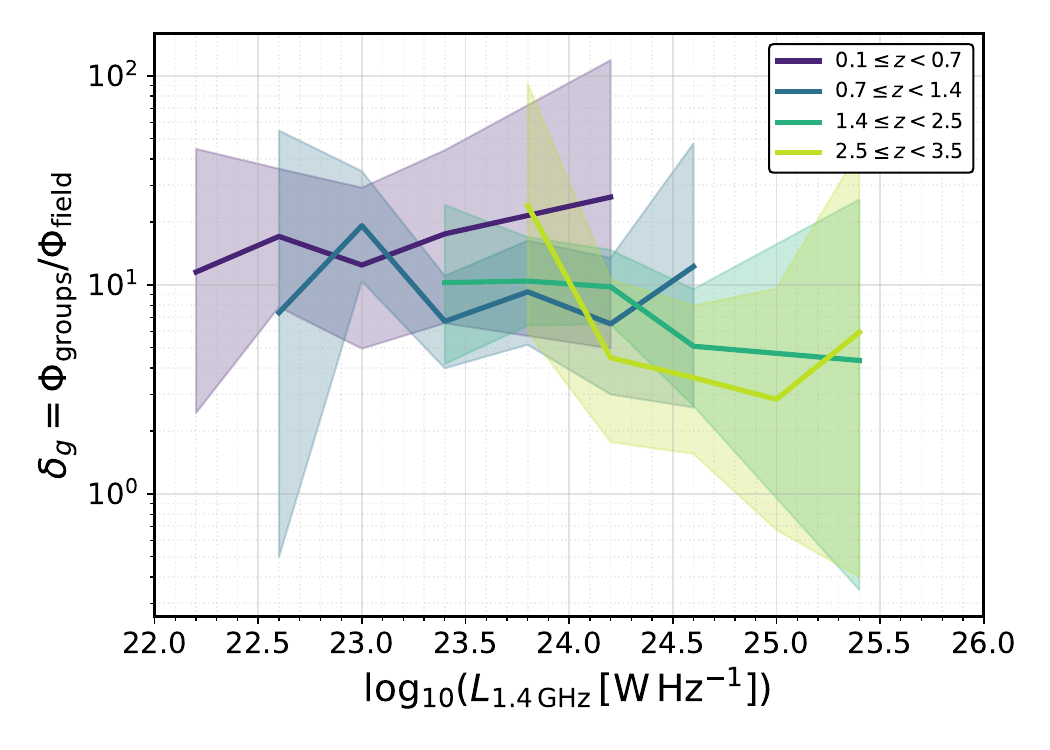}
        \includegraphics[width=0.47\linewidth]{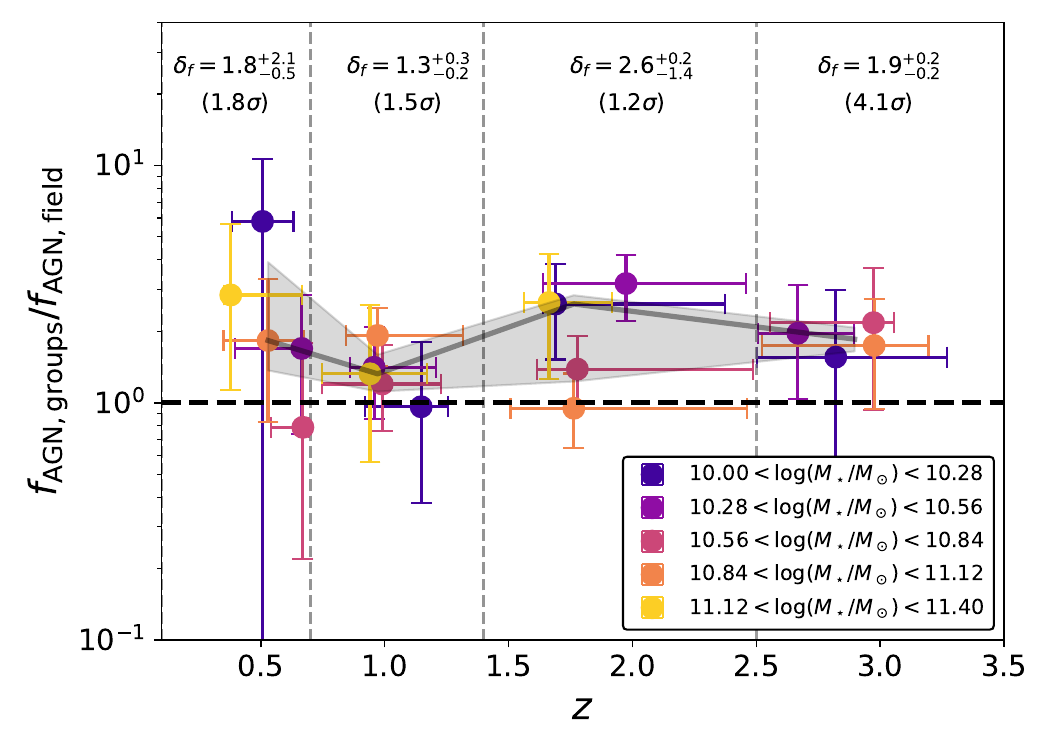}
    \caption{\textit{Left panel:} Ratio of the RLF of radio AGN in galaxy groups to the one of radio AGN in the field, as a function of radio luminosity and shown in bins of redshift. \textit{Right panel:} Redshift evolution of the ratio between the radio AGN fraction in galaxy groups and in the field (cumulative, for $L_{1.4}\geq10^{22}$~W~Hz$^{-1}$; from the points in Fig.~\ref{fig:fractions}), in bins of stellar mass (different colors). Points are plotted at the median $z$, with the uncertainties on the x-axis representing the 16$^{\rm th}$- and 84$^{\rm th}$-percentiles in each bin. The grey line (and shaded region) show the average ratio (and associated uncertainties) in the four redshift bins; the statistical significance of the excess with respect to $\delta_f = 1$ is reported in the upper part of the plot. These plots show that the group-to-field excess is a factor of $\delta_g\sim3-30$ in space density and a factor of $\delta_f\sim2$ (up to $4.1\sigma$ significance in the highest-$z$ bin) in AGN fraction.}\label{fig:excessinRLFandfractions}
\end{figure*}

\section{Discussion}\label{sec:disc}

\subsection{The importance of overdense environments in triggering radio AGN activity over cosmic time}\label{subsec:disc-overdensities}

\paragraph{Insights from the study of the RLF --}  Our investigation of the RLF, derived separately for radio AGN in groups and in the field (Sect.~\ref{subsub:buildRLF}, Sect.~\ref{subsub:fitRLF}), revealed a broadly similar shape of $\Phi (L)$ between the two environments, but a different normalization in the local Universe and a different evolution with redshift. Our fitting of the RLF (Sect.~\ref{subsub:fitRLF} and Table~\ref{tab:fitRLF}) allowed us to constrain, by extrapolation via our fitting function, the normalization of the local RLF for the two environments, $\log \Phi^{\ast} \equiv \log \Phi (L^{\ast},z=0)$. We found $\log \Phi^{\ast}{\rm [Mpc^{-3} dex^{-1}]}\rvert_{\rm groups}  = -4.82^{+0.58}_{-0.68}$ and $\log \Phi^{\ast}{\rm [Mpc^{-3} dex^{-1}]}\rvert_{\rm field}  = -5.29^{+0.19}_{-0.18}$, which implies a ratio $\delta_{g}^{z=0} = \frac{\Phi^{\ast}\rvert_{\rm groups}}{\Phi^{\ast}\rvert_{\rm field}} = 3.0^{+14.0}_{-2.5}$, i.e., an excess in the space density of radio AGN in groups relative to the field by approximately a factor of three. This excess persists up to high redshift. In Fig.~\ref{fig:excessinRLFandfractions} (left panel), we show the ratio between the observed RLF of radio AGN in galaxy groups and that in the field, $\delta_g = \Phi_{\rm groups}/\Phi_{\rm field}$, as a function of radio luminosity and in different redshift bins, using the RLFs derived in Sect.~\ref{subsub:buildRLF}; we find $\delta_g \approx 3-30$. 
The amplitude of the excess $\delta_{g}$ depends not only on the intrinsic overabundance of radio AGN in groups relative to the field, but also -- and most importantly -- on the adopted definition of the group volume (Sect.~\ref{subsub:buildRLF}). In this work, we defined the group volume as that enclosed within twice the root-mean-square projected group-centric galaxy distribution, corresponding to a characteristic radius of the order of $\approx$1~Mpc for the galaxy groups in the sample of \citet{Toni2025}. While adopting a different method to calculate the volume of groups would rescale $\Phi^{\ast}\rvert_{\rm groups}$, our choice of $R_{95\%}$ makes the inferred group space densities conservative. Since the adopted group volumes are likely upper limits (see Sect.~\ref{sec:methods}; for comparison, $R_{200}$ is on average $\sim30\%$ smaller than $R_{95\%}$), the measured excess of radio AGN in groups relative to the field can be seen as a conservative lower limit. An additional caveat is that our analysis is based on a single $\sim$0.45~deg$^{2}$ field, and as such, cosmic variance may be significant particularly at high-$z$. Confirmation of our trends using independent deep fields would represent an important follow-up of this work.
\\Our best-fit PDE models to the RLFs of radio AGN in groups and radio AGN in field (Sect.~\ref{subsub:fitRLF}) further seem to suggest, for the first time, that the number density of sources has a cosmic evolution (parametrized by the $\alpha_{D} + \beta_{D}z$ term) that proceeds slightly more rapidly in galaxy groups than in the field (Fig.~\ref{fig:RLFevo}). This is compatible with the evidence, based on Fig.~\ref{fig:RLF} and Fig.~\ref{fig:excessinRLFandfractions} (left panel), that $\delta_{g}$ decreases with increasing redshift.
\\\indent It is interesting to compare these results with the picture emerging from recent studies of X-ray selected AGN in overdense environments (high redshift clusters and proto-clusters). Our analysis shows that an AGN selection based on radio emission supports the idea that AGN activity is enhanced in overdense regions across cosmic time, as previously found for X-ray-selected AGN (e.g., \citealt{Tozzi2022,Travascio2025}). However, it is important to emphasize that the magnitude of the inferred enhancement depends strongly on the adopted AGN tracer (radio or X-ray) and the assumed volume estimate. For example, \citet{Traina2025}, studying X-ray selected AGN in three protoclusters at $z\sim2.2$ and $z\sim3.2$, reported an excess in the X-ray AGN space density relative to the field of a factor $\geq10\times - 100\times$. Similarly, \citet{Tozzi2022} found that the X-ray AGN luminosity function in the Spiderweb protocluster ($z\sim2.1$) is enhanced by a factor of $\sim5\times-20\times$ relative to the field. For comparison, our observed RLFs give $\delta_{g}\approx3-20$ at $1.4\leq z\leq3.5$ (Fig.~\ref{fig:excessinRLFandfractions}, left panel) and the best-fit PDE models to the RLF of radio AGN in groups and in the field (Table~\ref{tab:fitRLF}) give $\delta_{g}^{\rm fit} \simeq5 $ for $z=2.2$ and $\delta_{g}^{\rm fit} \simeq 2$ for $z=3.2$. At higher redshift, studies of X-ray AGN in protoclusters report even larger excesses. Studying the SPT2349-56 protocluster at $z=4.3$, \citet{Vito2024} found an AGN overdensity relative to the field of a factor $10^{2}\times-10^{4}\times$ depending on the assumed volume for the protocluster and the galaxy selection (see also \citealt{Travascio2025}). Therefore, at face values, it would seem that radio AGN are relatively less influenced by the environment than X-ray AGN at high redshift. However, this comparison should be taken with caution. First, the inferred AGN space density excess $\delta_{g}$ strongly depends on the adopted definition of the overdensity volume, which can vary substantially between analyses (see also \citealt{Vito2024,Traina2025}) and may also depend on the type of structure considered, namely groups and protocluster cores in this work versus massive protoclusters in studies such as \citet{Tozzi2022,Vito2024,Travascio2025,Traina2025}. Second, the overdensities hosting X-ray AGN reported at high redshift are individual structures, that might trace only the tip of the iceberg of overdense systems (see \citealt{IslaLLave2026arXiv} for a discussion of the challenges associated with comparing results inferred from individual protoclusters and AGN samples, selected through different methods). In contrast, our analysis adopts a population-based approach over a large sample of galaxy groups \citep{Toni2025} spanning a wide redshift range, thus tracing the average enhancement of radio AGN activity in dense environments rather than the properties of individual structures. A natural follow-up of this work is to perform the same analysis using X-ray selected AGN within COSMOS-Web and carry out a one-to-one comparison between radio- and X-ray-selected AGN in overdensities.

\paragraph{Insights from the study of radio AGN fractions --} The influence of the environment is also reflected in the radio AGN fractions. Using our cumulative radio AGN fractions (corresponding to $L_{1.4}\geq10^{22}$~W~Hz$^{-1}$; see Fig.~\ref{fig:fractions}), we show in Fig.~\ref{fig:excessinRLFandfractions} (right panel) the redshift evolution of the ratio between the cumulative radio AGN fractions in groups and in the field, in different stellar-mass bins. We observe that the ratio $\delta_f = f_{\rm AGN,\,groups}/f_{\rm AGN,\, field}$ varies in the range $1\lessapprox \delta_f\lessapprox6$. When averaging the different stellar mass bins, we find ratios of $\delta_f^{0.1\leq z<0.7} = 1.8^{+2.1}_{-0.5}$, $\delta_f^{0.7\leq z<1.4} = 1.3^{+0.3}_{-0.2}$, $\delta_f^{1.4\leq z<2.5} = 2.6^{+0.2}_{-1.4}$, and $\delta_f^{2.5\leq z<3.5} = 1.9^{+0.2}_{-0.2}$. These represent excess with respect to one with a significance of $1.8\sigma$, $1.5\sigma$, $1.2\sigma$, and $4.1\sigma$. We note that similar enhancements at $z\geq2$ were found by \citet{IslaLLave2026arXiv} for X-ray AGN in protoclusters.

We can also consider the best-fit of the radio AGN fractions further separated in bins of radio luminosity (Sect.~\ref{subsec:fitfractions}, Fig.~\ref{fig:fractionsall}, and Fig.~\ref{fig:bestfit-allfractions}). We note that, for $M_{\ast} = 10^{10}$~M$_{\odot}$, $L_{1.4} = 10^{22}$~W~Hz$^{-1}$, and $z=0$, our best fit models imply that the fraction of radio AGN in groups ($1.7_{-0.6}^{+2.8}\%$) is five times higher than the fraction of radio AGN in the field ($0.26_{-0.11}^{+0.19}\%$). This enhancement of the radio AGN fraction in the local Universe ($\delta_f^{z = 0} = 6.5^{+11.1}_{-5.3}$) is consistent with the excess in space density in the local Universe derived from the study of the RLF ($\simeq3$). At fixed stellar mass and radio luminosity, the redshift evolution of the radio AGN fraction also differs between the two environments. In the field, the radio AGN fraction increases monotonically with redshift, with a scaling that evolves from $f_{\rm AGN} ({\rm field} \rvert M_{\ast},L_{1.4},0\leq z \leq 1)\appropto(1+z)^{1.3}$ at $z \lesssim 1$ to a flatter dependence $f_{\rm AGN} ({\rm field} \rvert M_{\ast},L_{1.4},z \geq 3)\appropto(1+z)$ at higher redshift due to the $\delta z$ term (Table~\ref{tab:bestfit-fractions}). In contrast, the redshift evolution in galaxy groups is non-monotonic. The parametric form implies a decrease in the radio AGN fraction up to $z \sim 1.3$, followed by an increase at higher redshift. A similar effect can be seen in Figures 10 and 16 in \citet{Hashiguchi2023}, where the AGN fraction in clusters has a similar concave shape, with a minimum at $z\sim0.8$. This difference suggests that the physical mechanisms driving radio AGN activity may co-depend on redshift and environment. At high redshift, the increase in radio AGN fraction in both environments may be linked with the higher availability of cold gas and increased merger rates (e.g., \citealt{Eastman2007,Wang2017,Spingola2020,MAgliocchetti2025,MaioPeroux2026}). At lower redshift, the relative enhancement of AGN activity in groups might reflect the growing importance of a hot intragroup medium in sustaining radio-mode AGN activity (see also e.g., \citealt{Best2007,Sabater2013}). 

We note here that the fraction of SFG and QG contributing to the radio AGN population does not differ significantly between group and field environments, within the current statistical uncertainties. Therefore, with the available data, the classification of galaxies between SFG and QG does not seem to have a stronger influence on the radio AGN fraction than the environment (at fixed stellar mass and redshift). In this context, we note that \citet{Wang2024} finds that the probabilities of QG and SFG (in \textit{any} environment) of hosting a radio-excess AGN have a different evolution with stellar mass, radio luminosity (radio AGNs in QG are more likely associated with more massive and more radio luminous galaxies compared to SFG), and redshift (radio AGN fractions in SFG show a faster increase at high redshift than those in QG), but a very similar global normalization.

A further environmental ingredient in triggering radio AGN activity appears to be the location of galaxies within their host groups (Fig.~\ref{fig:fractions-radius}). Based on our analysis of Sect.~\ref{subsec:AGNfraction-vsradius}, we find that the radio AGN fraction decreases with increasing group-centric distance expressed as a fraction of $R_{200}$, especially for $z\leq1.4$. At earlier cosmic times, we observe a relative increase in the radio AGN fraction in the outskirts of the galaxy groups, for $1\,R_{200}\leq r\leq 3\,R_{200}$. We note here that \citet{Ghaffari2026} studied the radial dependence of the fraction of high luminosity AGN ($L_{X}\geq10^{44}$~erg~s$^{-1}$) in the 25 richest galaxy groups from \citet{Toni2025}, and similarly found that the AGN fraction within the central $0.2$~Mpc is higher than in the outskirts at $z\leq1$, while at $z > 1$ the
AGN fraction of the outskirts exceeds or match
the central values. One possible interpretation is that, at high redshift, we are observing groups during an active phase of mass assembly, with ongoing accretion of galaxies along filaments of the cosmic web. The elevated radio AGN fraction at large radii may therefore reflect enhanced triggering associated with these accreting structures (e.g., \citealt{Porter2008,Pimbblet2013,Umehata2019,deVos2024}; see also \citealt{Koulouridis2024} for a discussion of AGN triggering in the outskirts of -- local -- clusters of galaxies). Alternatively, \citet{Hashiguchi2023} suggested that the enhanced radio AGN fractions in the outskirts of galaxy clusters may be linked to cluster mergers, which could promote galaxy interactions and exchange of gas (see also \citealt{MAgliocchetti2025} for the effect of galaxy-galaxy mergers on radio AGN triggering). 


\begin{figure}[ht!]
    \centering
    \includegraphics[width=0.9\linewidth]{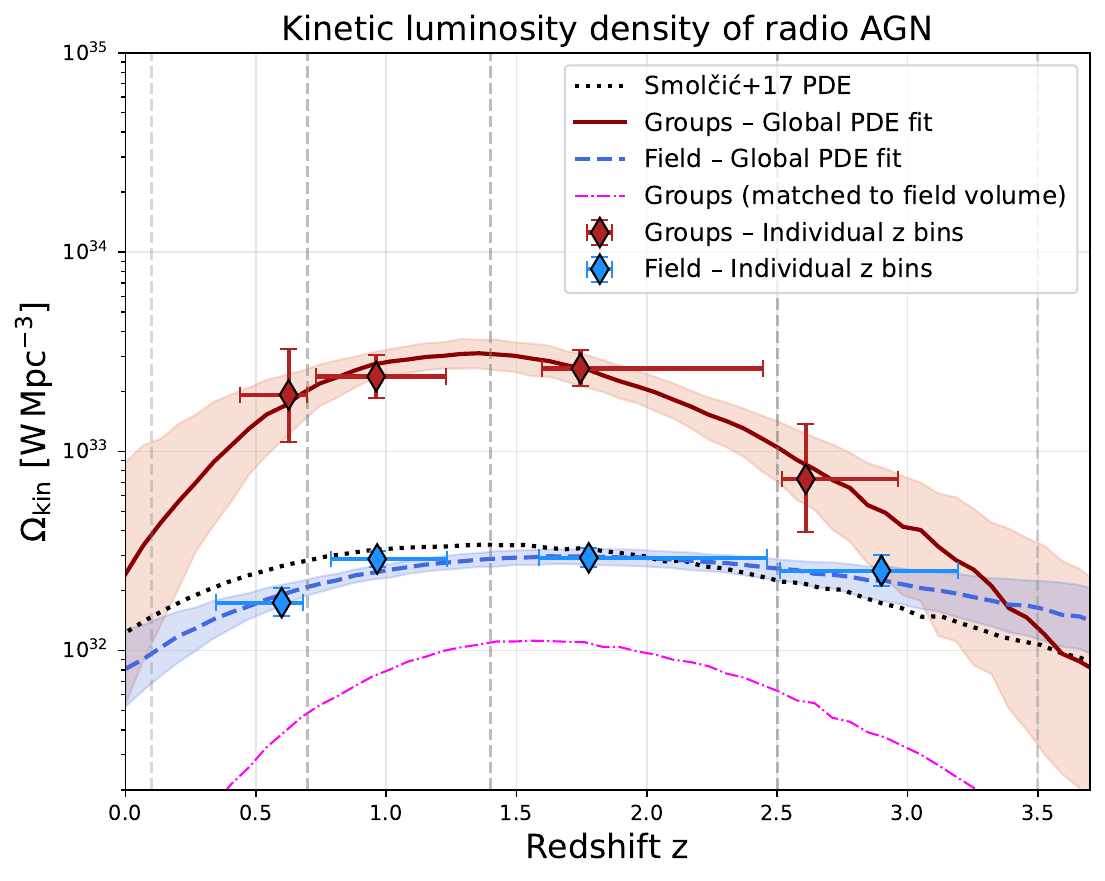}
    \caption{Cosmic evolution of the KLD of radio AGN in galaxy groups (red) and in the field (blue). Solid curves show the evolution inferred from the global PDE fits, while diamond symbols show the measurements obtained independently in each $z$ bin (Sect.~\ref{subsub:fitRLF}). The black dotted curves show the PDE evolution inferred from the global RLF of \citet{Smolcic2017_cosmicevo}. The dotted-dashed magenta line shows the KLD for radio AGN in groups obtained by using the \textit{same} volume as that used for radio AGN in the field, i.e., the $V_{\rm max}$ of each redshift bin (Eq.~\ref{eq:Vmaxfield}). The kinetic luminosity density is derived using the radio-to-kinetic conversion of \citet{Willott1999} with $f_{\rm W}=15$ (see Sect.~\ref{subsec:intRLF}).}
    \label{fig:integrRLF}
\end{figure}

\subsection{Feedback output of radio AGN in groups and in the field}\label{subsec:intRLF}

By integrating our best-fit RLFs separately for radio AGN in groups and in the field, we can derive the cumulative energy released over cosmic time -- specifically, the kinetic luminosity density (KLD).
The KLD is typically computed by assuming scaling relations from the literature (e.g., \citealt{Willott1999,Birzan2008,OSullivan2011,Kokotanekov2017,Grandi2021}) between the synchrotron radiative output of radio AGN (monochromatic or bolometric) and their mechanical energy output -- the latter is the amount of energy (or better, power) that is transferred to the medium. Assessing the cosmic evolution of the KLD of a population of radio AGN is a way to test how its global feedback energetics changed with $z$ (e.g., \citealt{Smolcic2017_cosmicevo,Delvecchio2022,Kondapally2022,Kondapally2023,Igo2025,Pierce2026}). The KLD is defined as:
\begin{equation}
    \label{eq:kindens}
    \Omega_{kin}(z) = \int P_{kin}(L_{1.4})\times \Phi(L_{1.4},z)\,d(\log L_{1.4})\, ,
\end{equation}
where $P_{kin}(L_{1.4})$ is the kinetic power corresponding to a monochromatic radio luminosity of $L_{1.4}$. The conversion between these two has been derived either from theoretical arguments on the efficiency of jet production in radio galaxies (e.g., \citealt{Willott1999}), or from the observation of X-ray cavities in clusters (e.g., \citealt{Birzan2008,OSullivan2011,Kokotanekov2017}), i.e., bubbles in the hot gas characterized by a low thermal gas density and a high relativistic gas density, excavated by relativistic jets while traveling through the intracluster medium (e.g., \citealt{Churazov2000,Birzan2004,Gitti2007,Blanton2011,Randall2015,David2017,Ubertosi2021} for some of the most studied systems). We adopted the formulation of \citet{Willott1999} (see also \citealt{Smolcic2017_cosmicevo}):
\begin{equation}
    \label{eq:willotpjet}
    \log P_{kin}(L_{1.4}) = 0.86 \cdot \log L_{1.4} + 14.08 + 1.5 \cdot \log f_{W}\, ,
\end{equation}
where $P_{kin}$ is in W, $L_{1.4}$ is in W~Hz$^{-1}$, and $f_{W}$ includes all the uncertainties in the scaling (mainly the jet composition, the energy production efficiency, and the minimum energy arguments, see \citealt{Willott1999} for a full discussion), and typically varies between 1 -- 20. We integrated the best-fit PDE models listed in Table~\ref{tab:fitRLF} using Equation~\ref{eq:kindens} and \ref{eq:willotpjet} between $10^{22}$~W~Hz$^{-1}$ and $10^{28}$~W~Hz$^{-1}$, as commonly done in the literature (e.g., \citealt{Smolcic2017_cosmicevo,Delvecchio2022}), and by assuming $f_{W} = 15$, which is the value that reproduces the correlation found from the study of X-ray cavities (see Appendix \ref{app:xstack} for a brief discussion of the implications of choosing a different $f_{W}$ for radio AGN in groups and in the field). We show the KLD for radio AGN in groups and in field in Fig.~\ref{fig:integrRLF}.

From Fig.~\ref{fig:integrRLF}, we can see that the 
KLD evolution of radio AGN in groups 
is
peaked at lower redshifts than that of radio AGN in field. In particular, it appears that the evolution of the radio AGN in groups population reaches its maximum at $1\leq z \leq 1.5$, while the corresponding peak for radio AGN in the field occurs at $z\sim2$. For comparison, in 
Fig.~\ref{fig:integrRLF} we also show the 
KLD of radio AGN in groups computed using the \textit{same} volume as that used for radio AGN in the field, i.e., the $V_{\rm max}$ of each redshift bin (Eq.~\ref{eq:Vmaxfield}). Since the volumes are equal in this case, the relative ratio of the integrated quantities between the two environments is driven simply by their relative numbers -- as already noted in Sect.~\ref{sec:methods}, radio AGN in the field account for 76\% of the total population. The purpose of including this additional line is thus to show that radio AGN in groups contribute $\sim1/4$ of the \textit{global} KLD of radio AGN in COSMOS shown in Fig.~\ref{fig:integrRLF} and in \citet{Smolcic2017_cosmicevo}. 
\\\indent Previous studies of the KLD evolution (e.g., \citealt{Smolcic2017_cosmicevo,Kondapally2023}) do not separate between the environment of radio AGN (see however \citealt{Pierce2026} for a discussion of its importance), thus, it is difficult to compare our results with other observational works. Nevertheless, we note that the KLD evolution derived for radio AGN in the field closely resembles that obtained for the global radio AGN population by \citet{Smolcic2017_cosmicevo}, which is itself consistent with several independent determinations of the KLD evolution (e.g., \citealt{Igo2025,Kondapally2023}). In contrast, the KLD of radio AGN in galaxy groups is systematically higher (reaching $\Omega_{\rm kin}\sim2$--$3\times10^{33}~{\rm W~Mpc^{-3}}$), implying that the cumulative mechanical energy released by SMBHs in dense environments exceeds that released in the field by roughly an order of magnitude for $0.5\lesssim z\lesssim2$. Most existing determinations of the KLD (\citealt{Kondapally2023} or \citealt{Pierce2026}; all derived for the \textit{global} radio AGN population, in field and in galaxy groups) lie below the values inferred here for radio AGN in groups (assuming the same $f_W$ conversion factor). 
The only comparable kinetic output we could identify is the theoretical prediction of \citet{Mocz2013} for a population composed of low-accretion-efficiency AGN with a kinetically-dominated feedback output. Their model predicts $\Omega_{\rm kin}\approx3$--$4\times10^{33}~{\rm W~Mpc^{-3}}$ over $0\leq z\leq3$ (their Fig.~5; see also Fig.~2 in \citealt{Kondapally2023}), in good agreement with the values of $\Omega_{\rm kin}$ we find for radio AGN in galaxy groups. For clarity, we note once again that the overall normalization of the observed $\Omega_{\rm kin}$ depends on the method chosen to compute the volume of groups. Therefore, despite the apparent agreement with the predictions of \citet{Mocz2013}, it is not straightforward to robustly conclude that the cumulative SMBH energy output in dense environments is dominated by low-accretion, jet-driven AGN feedback.

Ultimately, to estimate the potential impact of radio-mode feedback on the thermal history of group gas (see also \citealt{Hatch2014}), we integrated the median kinetic luminosity density of radio AGN in galaxy groups over $2\leq z<3.5$, since most present-day groups and clusters are expected to have assembled most of their mass by $z\sim2$ (e.g., \citealt{Overzier2016}). Adopting a cosmic baryon fraction $\Omega_{\rm b}=0.049$ and a hot gas fraction $f_{\rm g}=0.05
$ (e.g., \citealt{Lovisari2015}), we find that radio AGN heating can provide $\sim0.4$~keV per particle, consistent with the 
$0.1-1$~keV per particle typically invoked to pre-heat gas in galaxy groups \citep{Kaiser1991,McCarthy2010,Eckert2021}.

\section{Conclusions}\label{sec:summary}
We presented a multi-parametric study of radio AGN triggering in the COSMOS field, combining the VLA-COSMOS 3\,GHz Large Project with the COSMOS-Web galaxy and group catalogs to investigate the role of the large-scale environment in the redshift range $0.1\leq z\leq3.5$. Our results can be summarized as follows:

\begin{enumerate}[nolistsep]
    \item The RLF of radio AGN in galaxy groups is enhanced relative to that of radio AGN in the field by a factor $\delta_g \approx 3$--$30$, decreasing toward high $z$. The best-fit PDE parameters indicate a steeper cosmic evolution for radio AGN in groups than in the field, consistent with the observed decrease in $\delta_g$ with $z$. Although we choose a very conservative approach in deriving the maximum observable volume for radio AGN in galaxy groups, that maximizes the possible extent of the overdensities, we also highlight how different methods (using e.g., smaller radii for the galaxy groups) would directly impact the amplitude of the space density excess $\delta_g$.
    \item The radio AGN fraction in galaxy groups is, globally, higher than in the field at all redshifts ($0.1 \leq z \leq 3.5$), by a factor $\simeq2$. The radio AGN fraction increases with $M_{\ast}$ in both environments. The dependence on $L_{1.4}$ is flat in groups and mildly negative in the field, although the two slopes are consistent at $\sim$1.4$\sigma$. In the field, the fraction increases monotonically with $z$, while in galaxy groups, the fraction has a minimum at $z \sim 1.3$ and an increase at both lower and higher redshifts. For $z \lessapprox 1.5$, the radio AGN fraction increases toward group centers, with an enhancement of up to a factor of $\sim 4$ within $r \leq 0.2\,R_{200}$ relative to $r \gtrsim R_{200}$. At $z \gtrsim 1.4$, an elevated radio AGN fraction is visible up to $r \sim 3\,R_{200}$. 
    \item The KLD of radio AGN in groups seems to exceed that of radio AGN in the field by an order of magnitude (assuming the same $f_{W}$), and peaks at $1\lesssim z \lesssim 1.5$, compared to $z \simeq 2$ for the field population. This suggests that radio AGN in dense environments release the majority of their mechanical energy at later cosmic times than sources in the field.
    \item  Overall, we find that at $1\lesssim z \lesssim 1.5$ the radio AGN fraction in groups has a minimum and the radial distribution of radio AGN in groups changes. Notably, over the same redshift range radio AGN in groups also reach their peak space density. This may suggest that $1\lesssim z \lesssim 1.5$ represents a characteristic epoch in the co-evolution of radio AGN and their host structures.
\end{enumerate}

\begin{acknowledgements}
      FU, MG acknowledge the financial contribution from contract PRIN 2022 - CUP J53D23001610006. FU and ID acknowledge the financial support from the INAF project \say{U-JOIN}. CS acknowledges financial support by the Italian Ministry of University and Research (grant FIS2023$-$01611, CUP C53C25000300001) and by the INAF (through \textsl{Ricerca Fondamentale 2024}, Ob. Fu. 1.05.24.07.04).
\end{acknowledgements}

%
\bibliographystyle{aa} 
\bibliography{aa.bib} 

\begin{appendix}
\onecolumn

\section{The catalogs used in this work}
\begin{figure*}[htp!]
\centering
        \includegraphics[width=0.8\linewidth]{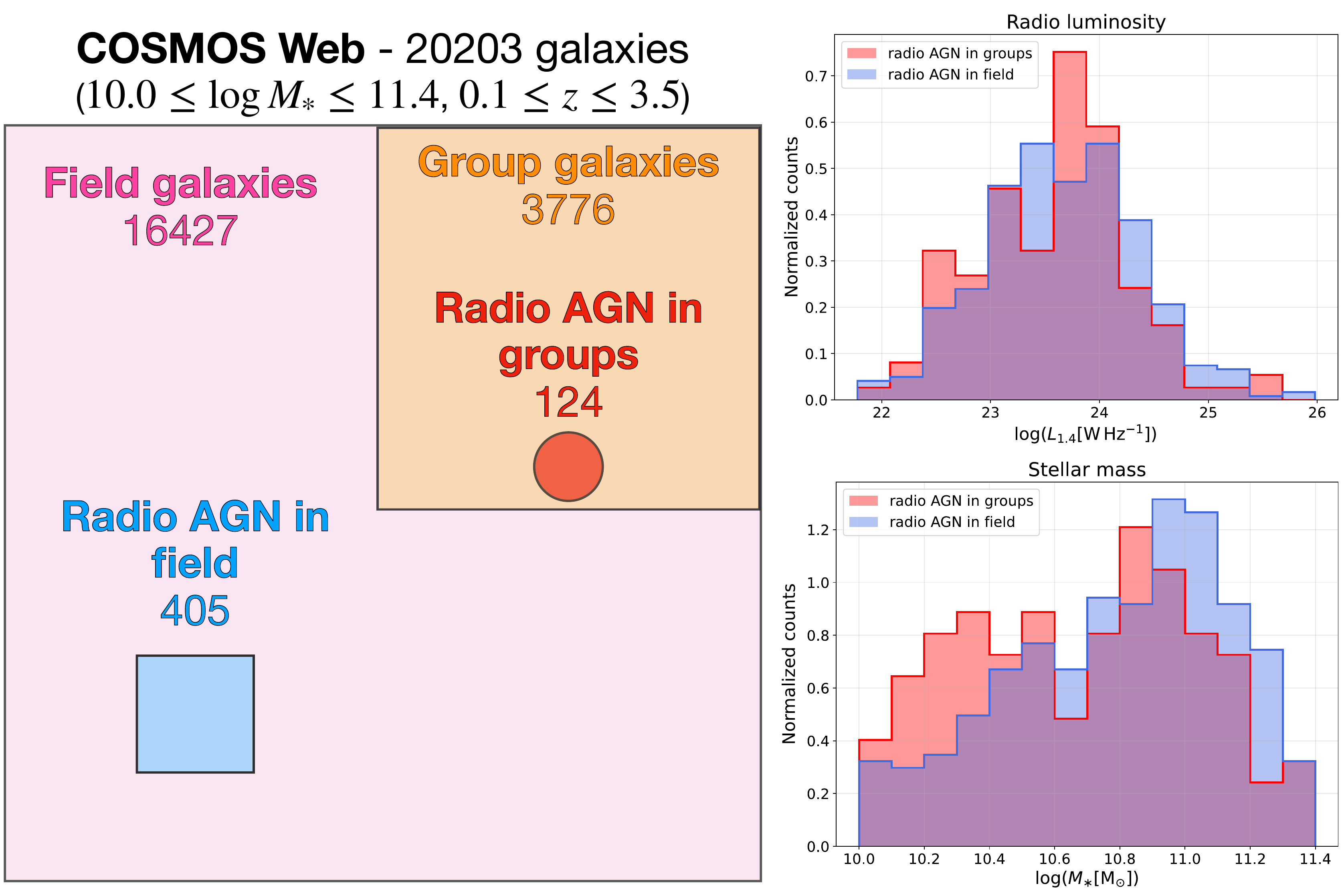}
        \caption{{\it Left:} Venn diagram of the catalogs used in this work. Number of objects are reported after mass ($10.0\leq\log({\rm M}_{\ast})\leq11.4$) and redshift ($0.1\leq z\leq 3.5$) cuts (see Sect.~\ref{subsec:createcat}). The areas of the different shapes are scaled by the number of objects they comprise. The total number of galaxies (20203) is the sum of all field and group galaxies (which already include the radio AGN sources). {\it Right:} Histograms (normalized by area) of radio luminosity $L_{1.4}$ (\textit{top}) and stellar mass $M_{\ast}$ (\textit{bottom}) for radio AGN in galaxy groups and radio AGN in field.}
        \label{fig:venn}
\end{figure*}

We provide in Fig.~\ref{fig:venn} a sketch of the different subsamples of galaxies (and corresponding catalogs) built as described in Sect.~\ref{subsec:createcat} and then analyzed in Sect.~\ref{sec:analysis} \& Sect.~\ref{sec:disc}. In the same Figure, we also show the radio luminosity and stellar mass distributions of radio AGN in galaxy groups and in the field.

\section{Fitting the RLF of radio AGN in the field and in galaxy groups}\label{app:fitRLF}
\begin{figure*}[htp!]
\centering
\includegraphics[width=0.49\textwidth]{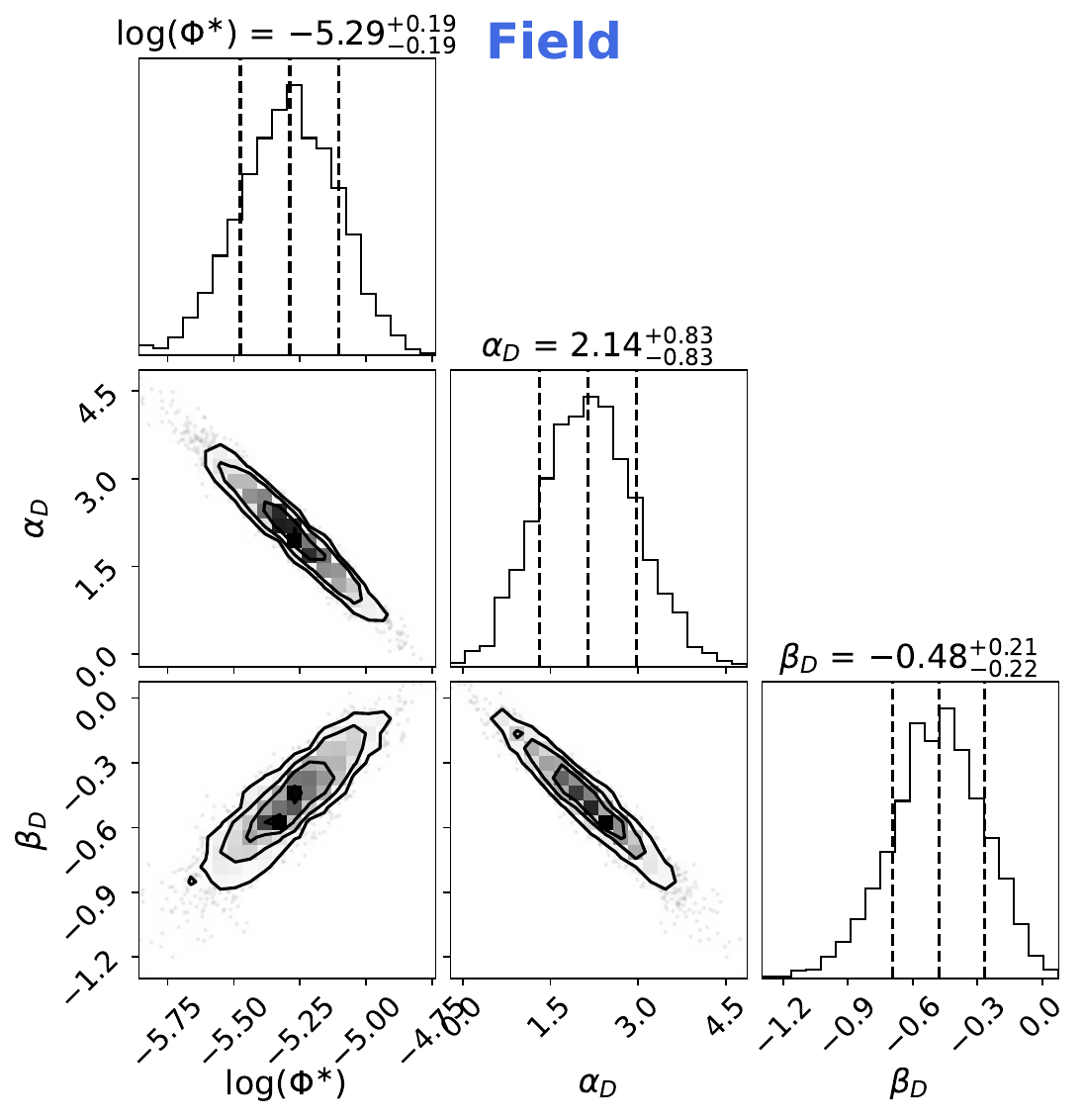}
\includegraphics[width=0.49\textwidth]{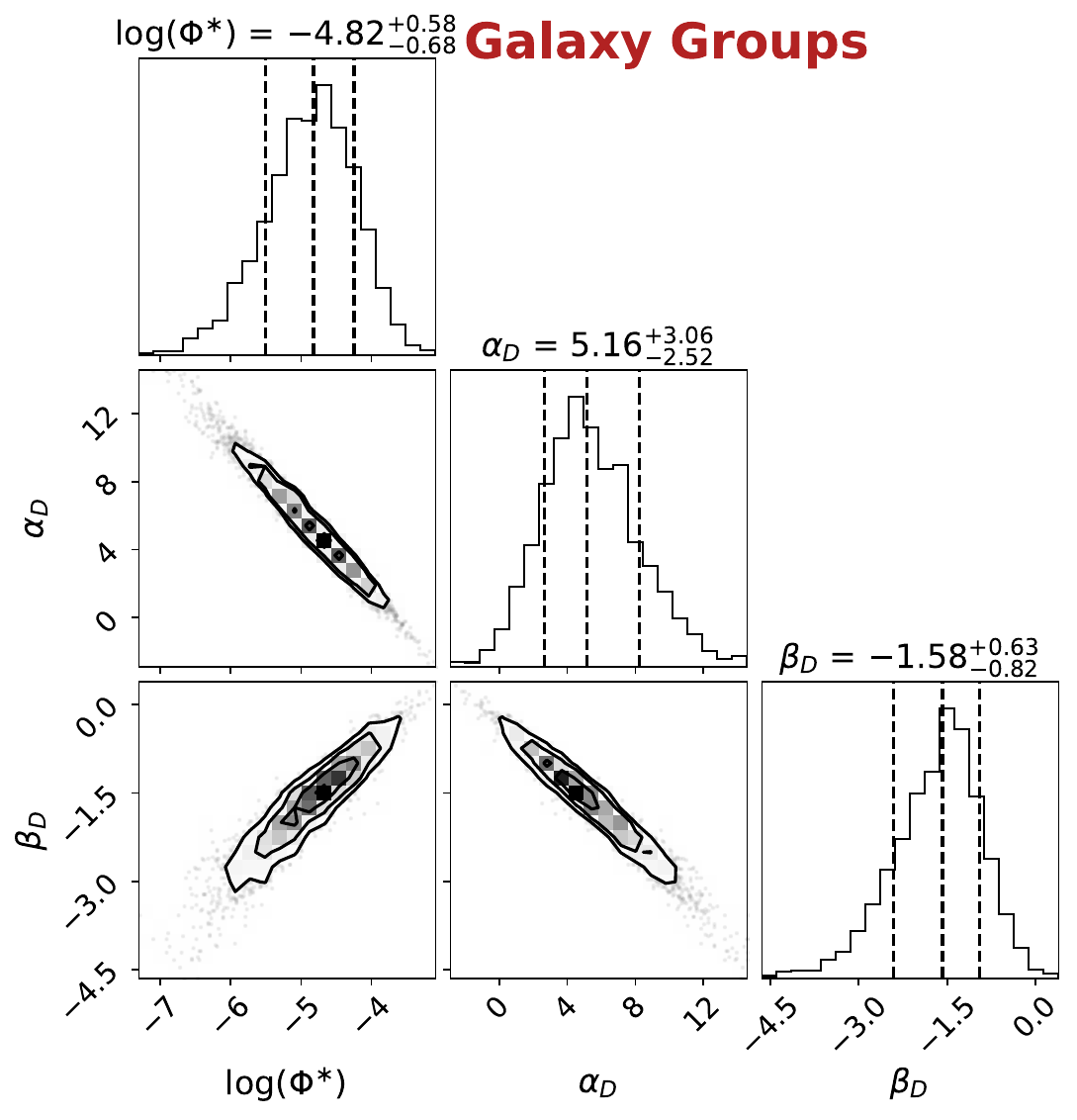}
\caption{Corner plots showing the posterior distributions of the fitted RLF parameters ($\Phi^{\ast}$, $\alpha_{D}$, $\beta_{D}$) for radio AGN in the field (left) and in groups (right). Contours represent the 68\%, 95\%, and 99.7\% confidence levels.}
\label{fig:cornerRLF}
\end{figure*}

In this Appendix, we show the corner plots associated to our fits of the RLF (Sect.~\ref{subsub:fitRLF}), performed separately for galaxies in galaxy groups and for galaxies in the field. Specifically, we show in Fig.~\ref{fig:cornerRLF} the posterior distributions of the parameters ($\Phi^{\ast}$, $\alpha_{D}$, $\beta_{D}$) fitted to the RLF for radio AGN in the field (left panel) and radio AGN in galaxy groups (right panel). We also report in Table~\ref{tab:fitRLF} the best-fit RLF parameters for radio AGN in different environments using a global PDE model (Eq.~\ref{eq:PDEcont}), and in Table~\ref{tab:fitRLF-zbins} the best-fit RLF parameters for radio AGN in different environments using a PDE model fitted to the individual redshift bins (Eq.~\ref{eq:PDE}).

\begin{table}[htp!]
\centering
\renewcommand{\arraystretch}{1.5}
\caption{Best-fit RLF parameters for radio AGN in different environments (see Sect.~\ref{subsub:fitRLF}).}
\label{tab:fitRLF}
\begin{tabular}{lcccc}
\hline
 & $\log \Phi^{\ast}$ & $\alpha_{D}$ & $\beta_{D}$ & Reduced $\chi^2$ \\
 & (Mpc$^{-3}$ dex$^{-1}$) &  &  &  \\
\hline
Field & $-5.29^{+0.19}_{-0.18}$ & $2.14^{+0.83}_{-0.83}$ & $-0.48^{+0.21}_{-0.22}$ & 24.9/22  \\
Groups & $-4.82^{+0.58}_{-0.68}$ & $5.16^{+3.06}_{-2.52}$ & $-1.58^{+0.63}_{-0.82}$ & 4.5/15 \\
\hline
\end{tabular}
\tablefoot{Columns: (1) environment; (2) normalization of the local RLF; (3) normalization of the pure-density global redshift evolution term in Eq.~\ref{eq:PDEcont}; (4) slope of the pure-density global redshift evolution term in Eq.~\ref{eq:PDEcont}; (5) Reduced $\chi^{2}$ of the best-fit.}
\end{table}

\begin{table}[htp!]
\centering
\renewcommand{\arraystretch}{2.0}
\caption{Best-fit RLF parameters (from the fit of the individual redshift bins) for radio AGN in different environments (see Sect.~\ref{subsub:fitRLF}).}
\label{tab:fitRLF-zbins}
\begin{tabular}{lcccc}
\hline
 & \multicolumn{2}{c}{Field} & \multicolumn{2}{c}{Groups} \\
\cline{2-3}\cline{4-5}
Redshift bin & $\alpha_{D}$ & $\chi^2_{\nu}$ & $\alpha_{D}$ & Reduced $\chi^2$ \\
\hline
$0.1 - 0.7$ & $1.68\pm0.35$ & $1.1/5$ & $4.43\pm1.11$ & $0.9/2$ \\
$0.7 - 1.4$ & $1.89\pm0.14$ & $7.8/5$ & $3.44\pm0.37$ & $0.8/3$ \\
$1.4 - 2.5$ & $1.27\pm0.11$ & $10.5/5$ & $2.39\pm0.21$ & $1.3/2$ \\
$2.5 - 3.5$ & $0.84\pm0.14$ & $3.2/2$ & $0.89\pm0.50$ & $1.1/3$ \\
\hline
\end{tabular}
\tablefoot{Columns: (1) redshift bin; (2) redshift evolution term of the PDE evolution model of Eq.~\ref{eq:PDE} for radio AGN in the field; (3) Reduced $\chi^{2}$ of the best-fit for radio AGN in the field; (4) redshift evolution term of the PDE evolution model of Eq.~\ref{eq:PDE} for radio AGN in galaxy groups; (5) Reduced $\chi^{2}$ of the best-fit for radio AGN in galaxy groups;}
\end{table}

\clearpage
\section{Plots and best-fit of radio AGN fractions as a function of M$_{\ast}$, redshift, $L_{1.4}$, and environment}\label{app:radioAGNfractions-parametric}
\begin{figure}[htp!]
    \centering
    \includegraphics[width=0.95\linewidth]{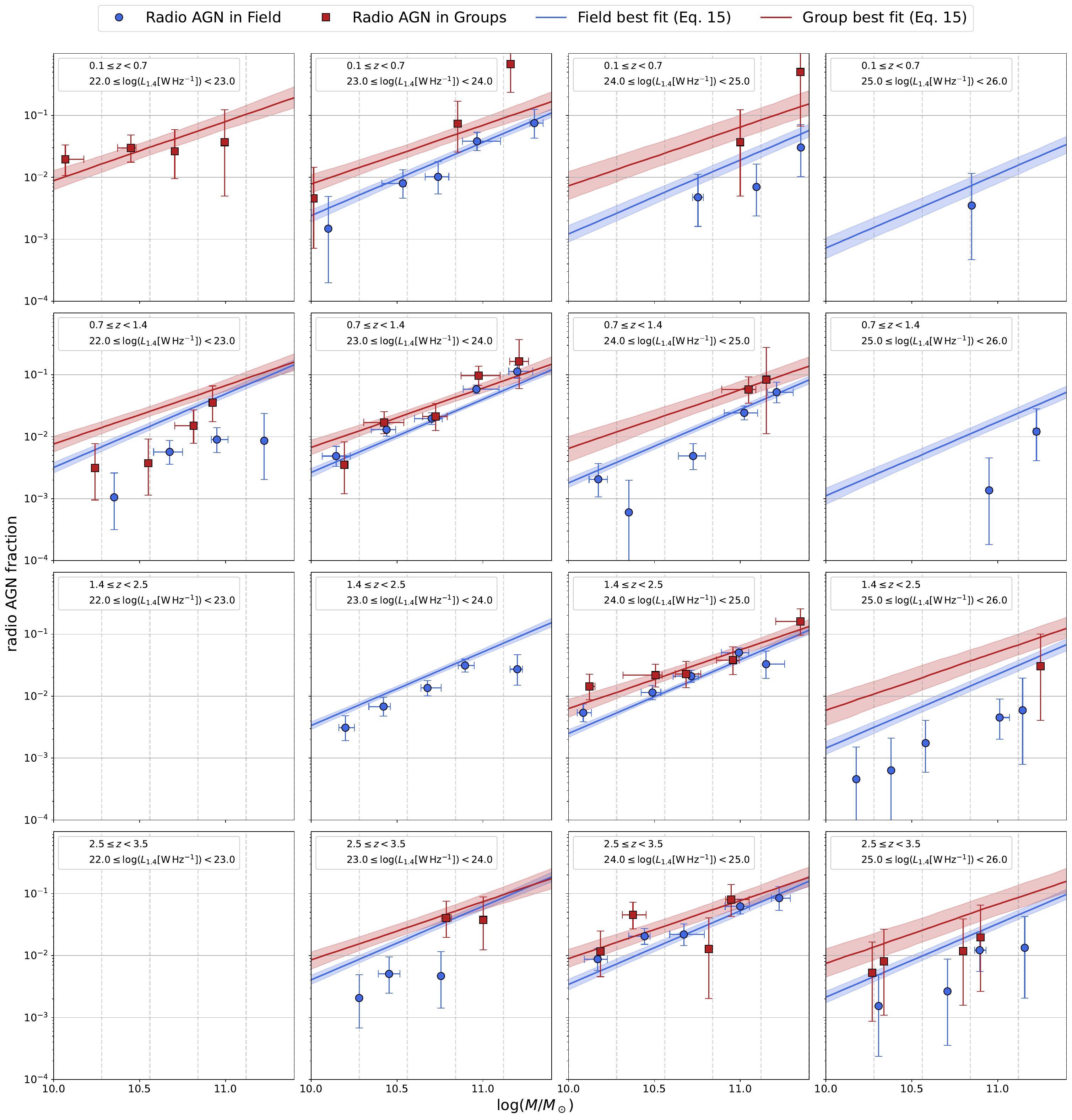}
    \caption{Radio AGN fraction as a function of stellar mass in four redshift bins between $0.1 \leq z \leq 3.5$ (from top to bottom) and in four luminosity bins of 1 dex width between $10^{22}~{\rm W\,Hz^{-1}} \leq L_{1.4} \leq 10^{26}~{\rm W\,Hz^{-1}}$ (from left to right). The best-fit parametric models and uncertainty interval from Sect.~\ref{subsec:fitfractions} (with parameters reported in Table~\ref{tab:bestfit-fractions}) are overlaid with solid lines and shaded areas, respectively, separately for the radio AGN in galaxy groups and in the field. Empty boxes show regions of the parameter space where we have no sources above the adopted completeness limit in radio luminosity.}
    \label{fig:fractionsall}
\end{figure}

\begin{figure*}[htp!]
\centering
\includegraphics[width=0.48\textwidth]{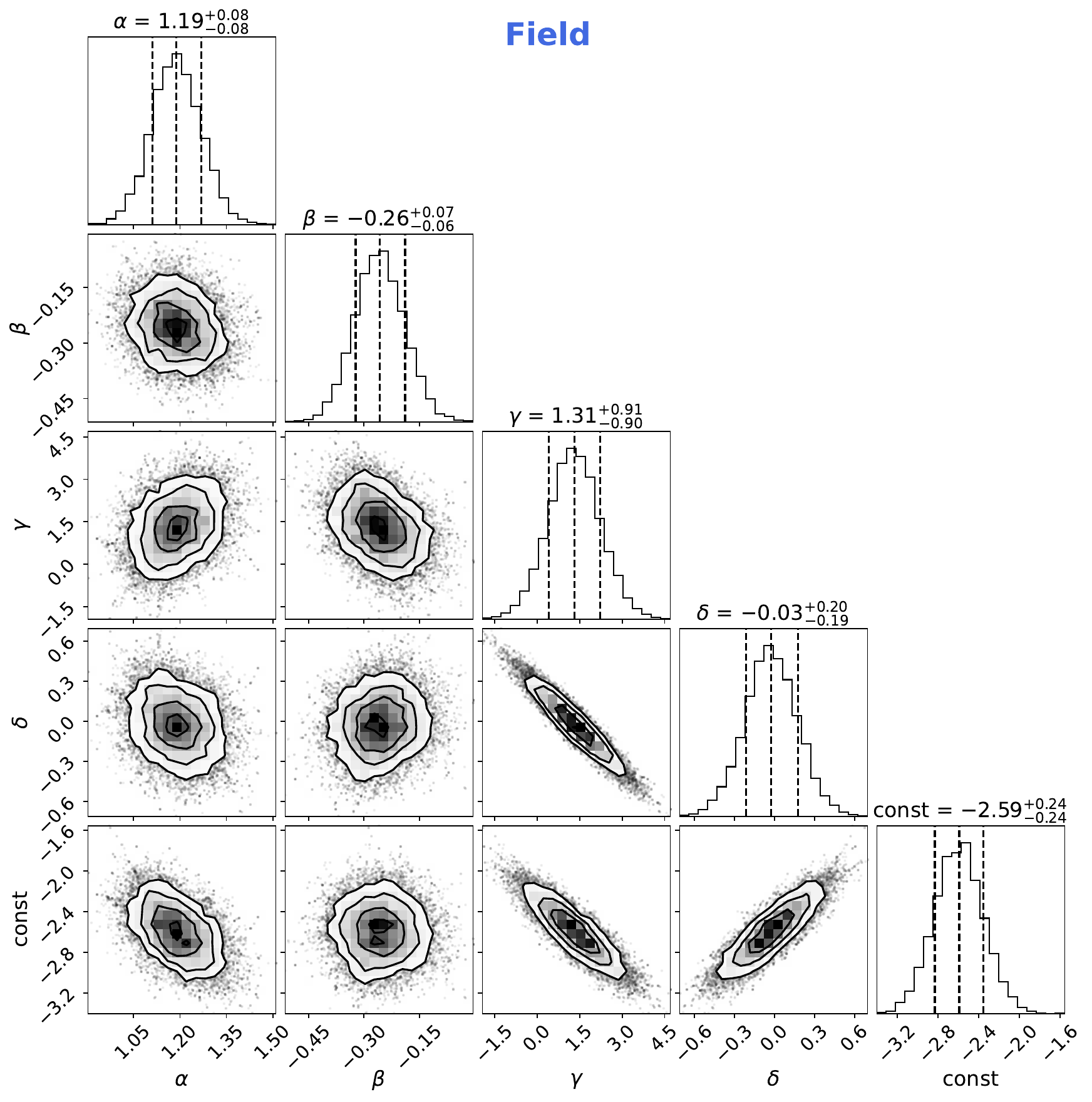}
\includegraphics[width=0.48\textwidth]{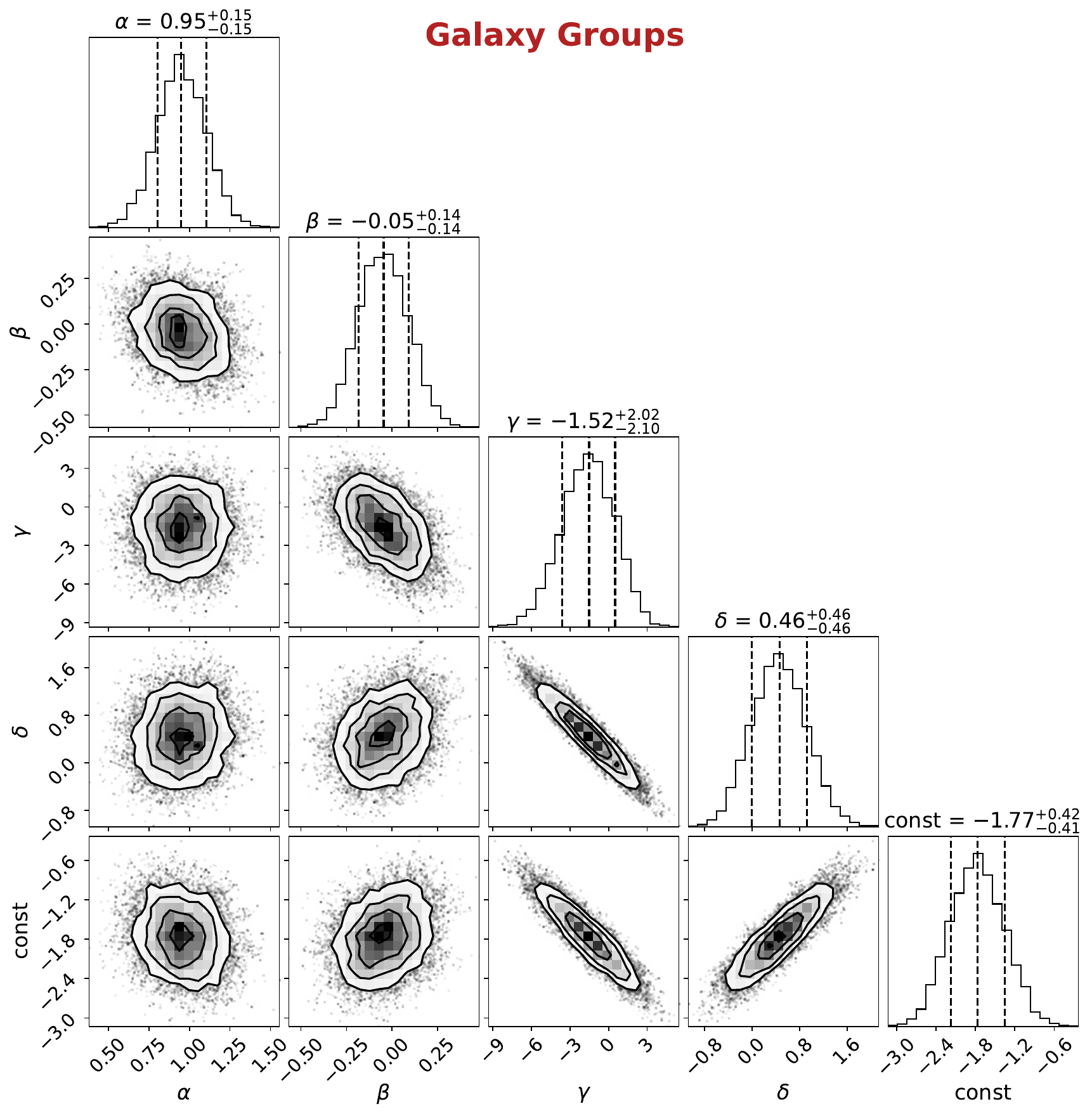}
\caption{Corner plots showing the posterior distributions of the fitted parameters to the radio AGN fractions (see Sect.~\ref{subsec:fitfractions}, Fig.~\ref{fig:fractionsall} and Fig.~\ref{fig:bestfit-allfractions}) for objects in the field (left) and in groups (right). The model is defined as 
$f_{\rm AGN} (M_{\ast},L_{1.4},z\,\rvert \,{\rm env})] = 10^{\rm const.}\left(\frac{M_{\ast}}{10^{10}\,\rm M_{\odot}}\right)^{\alpha}\left(\frac{L_{1.4}}{10^{22}\,\rm W~Hz^{-1}}\right)^{\beta}(1+z)^{(\gamma + \delta z)}$.
Contours represent the 68\%, 95\%, and 99.7\% confidence levels.}
\label{fig:corner-fractions}
\end{figure*}

In this Appendix, we present the plots of radio AGN fractions calculated separately for sources residing within galaxy groups and within the field, in bins of stellar mass, redshift and radio luminosity. Details are reported in Sect.~\ref{subsec:fractions}. In the plots of Fig.~\ref{fig:fractionsall}, we also overlay the best-fit parametric forms obtained in Sect.~\ref{subsec:fitfractions}. We also report in Table~\ref{tab:bestfit-fractions} the best-fit parameters and statistics of this fit, and we show in Fig.~\ref{fig:corner-fractions} the posterior distributions of the parameters ($\alpha,\,\beta,\,\gamma,\,\delta,\,{\rm const}$) fitted to the radio AGN fractions for galaxies in the field (left panel) and galaxies in galaxy groups (right panel). A table reporting the radio AGN fractions in separate galaxy environments, within the redshift, stellar mass, and radio luminosity bins shown in Fig.~\ref{fig:fractionsall}, and together with ancillary quantities, is available at this \href{https://github.com/FrancescoUbertosi/COSMOS-Radio-AGN-in-galaxy-groups.git}{Github link}.

\begin{table*}[htp!]
\centering
\renewcommand{\arraystretch}{2.0}
\caption{Best-fit parameters of the parametric fit to the fractions of radio AGN in different environments (see Sect.~\ref{subsec:fitfractions}, Fig.~\ref{fig:fractionsall} and Fig.~\ref{fig:bestfit-allfractions}). The model is defined as 
$f_{\rm AGN} (M_{\ast},L_{1.4},z\,\rvert \,{\rm env})] = 10^{\rm const.}\left(\frac{M_{\ast}}{10^{10}\,\rm M_{\odot}}\right)^{\alpha}\left(\frac{L_{1.4}}{10^{22}\,\rm W~Hz^{-1}}\right)^{\beta}(1+z)^{(\gamma + \delta z)}$ (see Eq.~\ref{eq:agnfractionfit}).
}
\label{tab:bestfit-fractions}
\begin{tabular}{lcccccc}
\hline
 & $\alpha$ & $\beta$ & $\gamma$ & $\delta$ & ${\rm const.}$& Reduced $\chi^2$ \\
\hline
Radio AGN in field   & $1.19^{+0.08}_{-0.08}$  & $-0.26^{+0.07}_{-0.06}$  & $1.31^{+0.91}_{-0.90}$  & $-0.03^{+0.20}_{-0.19}$  & $-2.59^{+0.24}_{-0.24}$  &   103.4/47 (2.20) \\
Radio AGN in galaxy groups   & $0.95^{+0.15}_{-0.15}$  & $-0.05^{+0.14}_{-0.14}$  &  $-1.52^{+2.02}_{-2.10}$ & $0.46^{+0.46}_{-0.46}$ & $-1.77^{+0.42}_{-0.41}$  &  28.4/31 (0.92)  \\
\hline
\end{tabular}
\tablefoot{Columns: (1) radio AGN population; (2) stellar-mass dependence; (3) 1.4 GHz radio luminosity dependence; (4) normalization of the redshift-evolution term; (5) slope of the redshift-evolution term; (6) normalization of the model; (7) Reduced $\chi^{2}$ of the best-fit.}
\end{table*}

\section{An alternative parametric form to fit the radio AGN fractions}
\label{app:fitfractions-complex}

As noted in Sect.~\ref{subsec:fitfractions}, our parametric fit to the radio AGN fractions of galaxies in the field does not provide a good description of the data in all $M_{\ast}$, $L_{1.4}$, and $z$ bins; specifically, the model tends to over-predict the data at high and low radio luminosities for a given redshift bin (Fig.~\ref{fig:fractionsall}). This suggests that a more complex model than a combination of power-laws may describe the radio AGN fractions, especially for sources in the field. Additionally, it is possible that the slope of radio AGN fractions with stellar mass has a redshift dependence that is not taken into account by the simple model of Equation~\ref{eq:agnfractionfit}, i.e., the slope might flatten at higher redshifts (see Fig.~\ref{fig:fractions}; see also \citealt{Wang2024}). Here we study whether the residual trends are due to co-dependencies between the three variables. We therefore adopt a second-order polynomial expansion of $\log(f_{\rm AGN})$ around the values $\log(M_{\ast}/{\rm M_{\odot}})=10$, $\log(L_{1.4}/{\rm W\,Hz^{-1}})=22$, and $\log(1+z)=0$. The new form is aimed at incorporating parabolic evolutions of radio AGN fractions in luminosity and redshift, together with cross-terms describing how the stellar-mass slope changes with radio luminosity and redshift, and how the luminosity slope evolves with redshift. The model is expressed as:

\begin{equation}
\begin{aligned}
\log[f_{\rm AGN}(M_{\ast},L_{1.4},z\,|\,{\rm env})] = &
A_0
+\alpha\,\Delta M
+\epsilon\,\Delta L
+\gamma\,\Delta Z 
+b_{L^2}\,\Delta L^2
+b_{Z^2}\,\Delta Z^2
+c_{ML}\,\Delta M\,\Delta L 
+c_{MZ}\,\Delta M\,\Delta Z
+\delta\,\Delta L\,\Delta Z ,\\
\quad {\rm where}\quad
\Delta M = &\log\left(\frac{M_{\ast}}{10^{10}\,{\rm M_{\odot}}}\right)\quad {\rm and} \quad
\Delta L=\log\left(\frac{L_{1.4}}{10^{22}\,{\rm W\,Hz^{-1}}}\right) \quad {\rm and} \quad
\Delta Z=\log(1+z).
\end{aligned}\label{eq:agnfractionfit-complex}
\end{equation}







\begin{figure}[ht!]
    \centering
    \includegraphics[width=\linewidth]{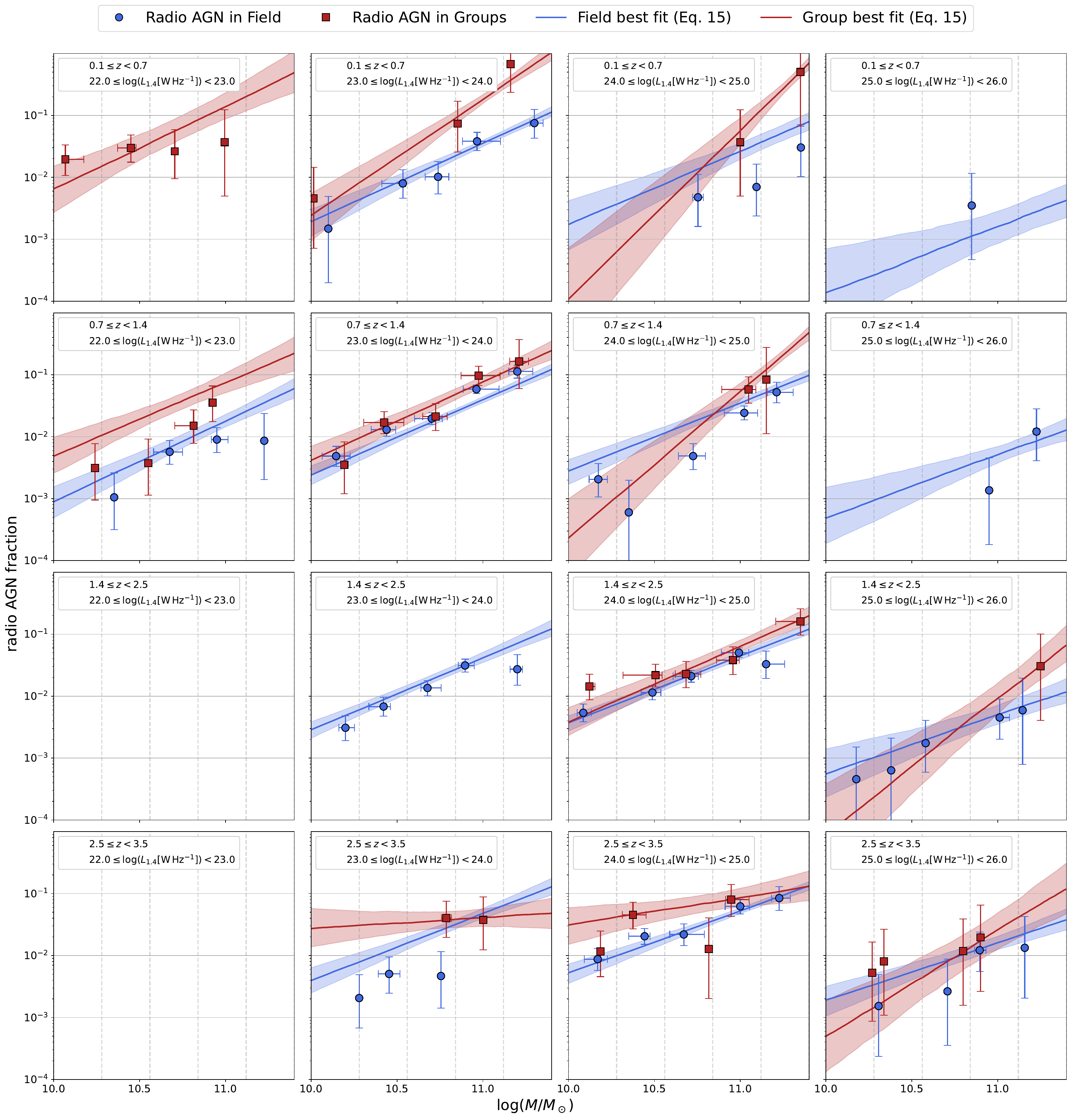}
    \caption{
    Analogous to Fig.~\ref{fig:fractionsall},
    but here the red and blue solid lines and shaded areas represent the more complex parametric model used to
    fit the radio AGN fractions for sources in the field, described in Appendix~\ref{app:fitfractions-complex}. This model has a parabolic dependence of $\log(f_{\rm AGN})$ on $L_{1.4}$ and $z$, as well as cross-dependencies between $M_{\ast}$, $L_{1.4}$, and $z$ (see Eq.~\ref{eq:agnfractionfit-complex} and Table~\ref{tab:bestfit-fractions-complex}). 
    We observe that this new model provides a better description of the radio AGN fractions, especially for sources in the field in the four $25\leq\log(L_{1.4})<26$ bins at all redshifts, and in the $22\leq\log(L_{1.4})<23$ bin at $0.7\leq z < 1.4$. However, some poorly fitted bins remain (the two $24\leq\log(L_{1.4})<25$ bins at $0.1\leq z < 0.7$ and $0.7\leq z < 1.4$, and the $23\leq\log(L_{1.4})<24$ bin at $2.5\leq z < 3.5$; see more in Appendix~\ref{app:fitfractions-complex}).
    }
    \label{fig:fractionsall-complex}
\end{figure}

This form allows both the normalization and the stellar-mass dependence of the radio AGN fractions to vary across radio luminosity and redshift. We fitted this model using the same methodology described in Sect.~\ref{subsec:fitfractions}. The resulting best-fit parameters are listed in Table~\ref{tab:bestfit-fractions-complex}, while the corresponding best-fit models are shown in Fig.~\ref{fig:fractionsall-complex}. Compared with the simpler model adopted in Sect.~\ref{subsec:fitfractions}, this model provides a lower reduced $\chi^2$, indicating that using higher-order terms and co-dependencies probably improves the parametric description of the radio AGN fractions (especially in the field). This is visible in Fig.~\ref{fig:fractionsall-complex}, where the best-fit model is now in better agreement with the field radio AGN fractions at both low and high radio luminosities. Nevertheless, some discrepancies remain in a few $(L_{1.4},z)$ bins, indicating that even this more complex form does not perfectly describe the data.

\begin{table*}[htp!]
\centering
\renewcommand{\arraystretch}{2.5}
\caption{
Best-fit parameters of the alternative parametric fit to the fractions of radio AGN in the field and in galaxy groups (see Appendix~\ref{app:fitfractions-complex} and Fig.~\ref{fig:fractionsall-complex}). The model is defined by Eq.~\ref{eq:agnfractionfit-complex}. 
}
\label{tab:bestfit-fractions-complex}
\resizebox{\textwidth}{!}{
\begin{tabular}{lcccccccccc}
\hline
 & $A_0$ & $\alpha$ & $\epsilon$ & $\gamma$ & $b_{L^2}$ & $b_{Z^2}$ & $c_{ML}$ & $c_{MZ}$ & $\delta$ & Reduced $\chi^2$ \\
\hline

Radio AGN in field &
$-3.85^{+0.57}_{-0.59}$ &
$1.48^{+0.39}_{-0.40}$ &
$1.69^{+0.49}_{-0.48}$ &
$-2.29^{+2.09}_{-2.01}$ &
$-0.62^{+0.13}_{-0.13}$ &
$-1.54^{+2.15}_{-2.26}$ &
$-0.12^{+0.29}_{-0.31}$ &
$-0.29^{+1.15}_{-1.10}$ &
$2.10^{+0.80}_{-0.77}$ &
58.5/43 (1.36) \\

Radio AGN in groups &
$-1.61^{+0.83}_{-0.81}$ &
$2.20^{+0.68}_{-0.67}$ &
$-0.42^{+0.73}_{-0.71}$ &
$-3.80^{+4.28}_{-4.27}$ &
$-0.71^{+0.22}_{-0.23}$ &
$-1.09^{+5.17}_{-4.91}$ &
$1.00^{+0.49}_{-0.50}$ &
$-7.14^{+2.44}_{-2.49}$ &
$5.60^{+1.98}_{-2.01}$ &
27.8/27 (1.03) \\
\hline
\end{tabular}
}
\tablefoot{Columns: (1) radio AGN population; (2) normalization of the model in Eq.~\ref{eq:agnfractionfit-complex}; (3) stellar-mass dependence; (4) 1.4 GHz radio luminosity dependence; (5) redshift dependence; (6) quadratic 1.4 GHz radio luminosity dependence; (6) quadratic redshift dependence; (7) stellar mass - radio luminosity cross dependence; (8) stellar mass - redshift cross dependence; (9) redshift - 1.4 GHz radio luminosity cross dependence; (10) reduced $\chi^{2}$ of the best-fit.}
\end{table*}

We stress that this new model aims at providing a more complex, but also more flexible, description of the observed trends and should not be interpreted as a physically motivated model for explaining the occurrence of radio AGN. Moreover, the larger number of free parameters makes the comparisons between the best-fit models for the two different environments less straightforward than with the simpler power-law model adopted in Sect.~\ref{subsec:fitfractions}. For these reasons, throughout the main discussion we retain the simpler parametrization, which provides a more straightforward description of the dependence of the radio AGN fraction on stellar mass, radio luminosity, and redshift, while this Appendix is only meant to demonstrate that some of the remaining residuals may arise from co-dependencies between these quantities.

\section{X-ray stacking of radio-excess AGN in groups and in field environments}\label{app:xstack}
The inferred normalization of the KLD derived in Sect.~\ref{subsec:intRLF} strongly depends on the conversion between radio luminosity and jet kinetic power. In this work, we assumed the same value of $f_W=15$ for radio AGN in groups and in the field. Physically, however, the effective value of $f_W$ may depend on environment, if the coupling between SMBH accretion and jet production is itself influenced by the dense environments in which these AGN reside. Systems accreting at low Eddington ratios are expected to have a larger fraction of their energy output being put into mechanical jet power rather than radiative power (e.g., \citealt{Nagar2005}; see also \citealt{Grandi2021}). Since highly-accreting AGN are rarely found in dense environments in the local Universe \citep{Donahue2020}, the true radio-to-kinetic conversion factor could differ between radio AGN in groups and in the field (e.g., \citealt{Igo2024}). In particular, if radio AGN in groups systematically accrete at lower Eddington ratios than those in the field, they may correspond to larger effective $f_W$ values and therefore larger kinetic luminosity densities than inferred here. At the same time, however, the fraction of highly accreting AGN in dense environments increases with $z$ (e.g., \citealt{HlavacekLarrondo2013}; see also \citet{Bufanda2017} for a study of increasing X-ray AGN fraction with $z$), so that beyond the local Universe, highly-accreting AGN would be favored \citep{McDonald2021}. Thus, differences in $f_W$ for radio AGN in different environments might disappear at $z \gtrapprox1$. 

To explore this possibility, we performed an X-ray stacking analysis of radio AGN in groups and radio AGN in the field, to check if radio AGN in field have relatively higher X-ray luminosities than radio AGN in galaxy groups. For X-ray stacking we used the publicly available Chandra tool \href{http://lambic.astrosen.unam.mx/cstack/}{CSTACK} developed by T. Miyaji. For each redshift bin and environment (field vs group), we stacked the soft (0.5-2) and hard (2-8 keV) images combining both X-ray detections and non-detections altogether. We verified that stacking non-detections and averaging them a-posteriori to the mean flux of X-ray detections does not change the results, though the final stacks are largely dominated by individual detections. From the soft and hard-band counts, we estimated the hardness ratio and a correction for nuclear obscuration (see e.g. \citealt{Xue2010}), by assuming an intrinsic power-law X-ray spectrum with constant slope $\Gamma=1.8$ (e.g. \citealt{Tozzi2006}). Finally, we used the de-obscured hard-band flux to compute the rest-frame 2-10 keV luminosity (or its upper limit in the highest $z$-bin for groups). Whenever measured, the $L_{X}$ values are always much higher than expected from star formation \citep{Lehmer2016}, hence no formal SF subtraction was done. The intrinsic 2-10 keV luminosities were scaled to bolometric via a single bolometric correction of 25 (e.g., \citealt{Aird2018}), and then converted to Eddington ratios by assuming a constant M$_{\rm BH}$/M$_{\ast}$ (black hole to bulge mass) of 0.002 (e.g. \citealt{Marconi2003}). Overall, we do not find significant differences in $L_{X}$ or Eddington ratio between field and group stacks, at fixed redshift, although we acknowledge the limited statistics (especially for radio AGN in groups), which makes these results dominated by the few X-ray detections. A wider survey would be needed to obtain more radio AGN and galaxy groups and understand if these X-ray stacking results are limited by the statistics or not. Future or ongoing spectroscopic surveys (e.g., COSMOS-3D, \citealt{Kakiichi2022}) might improve the group selection. In Sect.~\ref{subsec:intRLF}, we proceed with our -- potentially simplifying -- assumption of a common $f_W$, with the goal to perform a fair comparison between radio AGN in different environments.
\end{appendix}
\end{document}